\newif\iflinenums
\linenumsfalse
\def\docopts{preprint}
\def\docclass{aastex}
\def\hypcap{all}
\def\authlist{authors.tex}
\def\tabletype{deluxetable}

\ifdefined\des
  \linenumsfalse
\fi

\ifdefined\aastex
  \linenumsfalse
\fi

\ifdefined\arxiv
  \linenumsfalse
  \def\docopts{iop,revtex4-1}
  \def\docclass{hackemulateapj}
  \def\authlist{authors_revtex.tex}
  \def\hypcap{figure,figure*}
  \def\tabletype{deluxetable*}
\fi

\documentclass[\docopts]{\docclass}

\usepackage{graphicx}
\usepackage[dvipsnames,svgnames]{xcolor} 
\usepackage{natbib} 
\usepackage{hyperref}
\usepackage{amsmath}
\usepackage{multirow}
\usepackage{accents}
\hypersetup{    
  colorlinks      = true,
  linkcolor       = {black},
  linkbordercolor = {white},
  citecolor       = {black},
  citebordercolor = {white},
  urlcolor        = {blue},
  urlbordercolor  = {white},
}
\usepackage[\hypcap]{hypcap}

\usepackage{soul} 
\usepackage{amsmath}
\usepackage{xspace}
\usepackage{lineno}

\graphicspath{{./}{./figs/}}

\usepackage{soul} 
\usepackage{amsmath}
\usepackage{xspace}
\usepackage{xifthen}

\newcommand{\ie}{i.e.\xspace}
\newcommand{\eg}{e.g.\xspace}
\newcommand{\etc}{etc.\xspace}

\newcommand{\FIXME}[1]{{#1}}
\newcommand{\CHECK}[1]{{#1}}

\newcommand{\NEW}[1]{}

\mathchardef\mhyphen="2D

\newcommand{\roughly}{\ensuremath{ {\sim}\,} }

\newlength{\dhatheight}

\newcommand{\code}[1]{\texttt{#1}\xspace}

\newcommand{\var}[1]{\ensuremath{#1}\xspace}

\newcommand{\unit}[1]{\ensuremath{\mathrm{\,#1}}\xspace}
\newcommand{\Gyr}{\unit{Gyr}}

\newcommand{\degree}{\ensuremath{{}^{\circ}}\xspace}
\newcommand{\mas}{\unit{mas}}

\newcommand{\asec}{\unit{arcsec}}

\newcommand{\pc}{\unit{pc}}
\newcommand{\kpc}{\unit{kpc}}

\newcommand{\Msun}{\ensuremath{M_\odot}}
\newcommand{\Mstar}{\ensuremath{M_{*}}}

\newcommand{\magn}{\unit{mag}}
\newcommand{\mmag}{\unit{mmag}}
\renewcommand\ion[2]{#1$\;${\scshape{#2}}}

\newcommand{\secref}[1]{Section~\ref{sec:#1}}

\newcommand{\tabref}[1]{Table~\ref{tab:#1}}

\newcommand{\figref}[1]{Figure~\ref{fig:#1}}

\newcommand{\Nobjs}{\FIXME{Six}\xspace}
\newcommand{\nobjs}{\FIXME{six}\xspace}
\newcommand{\Ncand}{\FIXME{Two}\xspace}
\newcommand{\ncand}{\FIXME{two}\xspace}

\newcommand{\ntotal}{\FIXME{eight}\xspace}

\newcommand{\nyearI}{\FIXME{nine}\xspace}

\newcommand{\nsys}{\FIXME{eleven}\xspace}

\newcommand{\ndsphs}{\FIXME{fourteen}\xspace}

\newcommand{\ndes}{\FIXME{seventeen}\xspace}

\newcommand{\nsdss}{\FIXME{sixteen}\xspace}

\newcommand{\interval}{\FIXME{68\%}\xspace}
\newcommand{\limit}{\FIXME{84\%}\xspace}

\newcommand{\lummax}{{-4.7}\xspace}
\newcommand{\distmin}{{25}\xspace}
\newcommand{\distmax}{{214}\xspace}
\newcommand{\sizemin}{{17}\xspace}
\newcommand{\sizemax}{{181}\xspace}

\newcommand{\kimtwo}{{DES\,J2108.8\allowbreak$-$5109}\xspace}
\newcommand{\luque}{{DES\,1}\xspace}

\newcommand{\eriII}{{DES\,J0344.3\allowbreak$-$4331}\xspace}
\newcommand{\tucII}{{DES\,J2251.2\allowbreak$-$5836}\xspace}

\newcommand{\picI}{{DES\,J0443.8\allowbreak$-$5017}\xspace}
\newcommand{\eriIII}{{DES\,J0222.7\allowbreak$-$5217}\xspace}
\newcommand{\pheII}{{DES\,J2339.9\allowbreak$-$5424}\xspace}

\newcommand{\colI}{{DES\,J0531\allowbreak$-$2801}\xspace}
\newcommand{\cetII}{{DES\,J0117\allowbreak$-$1725}\xspace}
\newcommand{\gruII}{{DES\,J2204\allowbreak$-$4626}\xspace}
\newcommand{\indII}{{DES\,J2038\allowbreak$-$4609}\xspace}
\newcommand{\retIII}{{DES\,J0345\allowbreak$-$6026}\xspace}
\newcommand{\tucIII}{{DES\,J2356\allowbreak$-$5935}\xspace}
\newcommand{\tucIV}{{DES\,J0002\allowbreak$-$6051}\xspace}
\newcommand{\tucV}{{DES\,J2337\allowbreak$-$6316}\xspace}

\newcommand{\ngc}[1]{{NGC\,#1}\xspace}

\newcommand{\SExtractor}{\code{SExtractor}}
\newcommand{\HEALPix}{\code{HEALPix}}

\newcommand{\emcee}{\code{emcee}}

\newcommand{\bandvar}[2][]{%
  \ifthenelse{\isempty{#1}}{\var{#2}}{\var{#2\_#1}}%
}
\newcommand{\spreadmodel}[1][]{\bandvar[#1]{spread\_model}}
\newcommand{\spreaderrmodel}[1][]{\bandvar[#1]{spreaderr\_model}}
\newcommand{\wavgspreadmodel}[1][]{\bandvar[#1]{wavg\_spread\_model}}
\newcommand{\classstar}[1][]{\bandvar[#1]{class\_star}}

\newcommand{\magerrpsf}[1][]{\bandvar[#1]{magerr\_psf}}
\newcommand{\flags}[1][]{\bandvar[#1]{flags}}

\newcommand{\modulus}{\ensuremath{m - M}\xspace}
\newcommand{\ra}{{\ensuremath{\alpha_{2000}}}\xspace}
\newcommand{\dec}{{\ensuremath{\delta_{2000}}}\xspace}
\newcommand{\age}{{\ensuremath{\tau}}\xspace}
\newcommand{\metal}{{\ensuremath{Z}}\xspace}

\newcommand{\TS}{\ensuremath{\mathrm{TS}}\xspace}

\providecommand\physrep{\ref@jnl{Phys.~Rep.}}%
\providecommand\apjs{\ref@jnl{ApJS}}%
\providecommand{\jcap}{\ref@jnl{JCAP}}%
\begin{document} 

\iflinenums
  \linenumbers
\fi

\title{Eight Ultra-faint Galaxy Candidates Discovered in Year Two of the Dark Energy Survey} 
\input{\authlist}
\date{}

%TC:break Abstract
\begin{abstract}

We report the discovery of \ntotal new ultra-faint dwarf galaxy candidates in the second year of optical imaging data from the Dark Energy Survey (DES).
\Nobjs of these candidates are detected at high confidence, while \ncand lower-confidence candidates are identified in regions of non-uniform survey coverage.
The new stellar systems are found by three independent automated search techniques and are identified as overdensities of stars, consistent with the isochrone and luminosity function of an old and metal-poor simple stellar population.
The new systems are faint (\CHECK{$M_V > \lummax \magn$}) and span a range of physical sizes (\CHECK{$\sizemin \pc < r_{1/2} < \sizemax \pc$}) and heliocentric distances (\CHECK{$\distmin\kpc < D_\odot < \distmax\kpc$}).
All of the new systems have central surface brightnesses consistent with known ultra-faint dwarf galaxies (\CHECK{$\mu \gtrsim 27.5 \magn \asec^{-2}$}).
Roughly half of the DES candidates are more distant, less luminous, and/or have lower surface brightnesses than previously known Milky Way satellite galaxies.
%, and would have had a low probability of detection if observed by the Sloan Digital Sky Survey. 
Most of the candidates are found in the southern part of the DES footprint close to the Magellanic Clouds. 
We find that the DES data alone exclude ($p < 10^{-3}$) a spatially isotropic distribution of Milky Way satellites and that the observed distribution can be well, though not uniquely, described by an association between several of the DES satellites and the Magellanic system.
Our model predicts that the full sky may hold $\roughly100$ ultra-faint galaxies with physical properties comparable to the DES satellites and that 20--30\% of these would be spatially associated with the Magellanic Clouds.

%\FIXME{Abstract can be at most 250 words.}

\keywords{galaxies: dwarf --- galaxies: Local Group}
%\pacs{95.35.+d, 95.85.Pw, 98.52.Wz}
\end{abstract}
%TC:break _main_

\maketitle 

\section{Introduction}
\label{sec:intro}

%%% Science-focused paragraph %%%

The population of Milky Way satellite galaxies includes the least luminous, least chemically enriched, and most dark matter dominated galaxies in the known Universe \citep[\eg,][]{2007ApJ...670..313S,Kirby:2008}.
Although extreme systems from an observational perspective, low-luminosity dwarf spheroidal galaxies are likely to be the most common galaxy type by number.
This duality places the emerging population of near-field galaxies in a unique position to test models of galaxy formation and the cold dark matter paradigm \citep[\eg,][]{Bullock:2001}, and has motivated studies ranging from the physical conditions at the time of reionization \citep[\eg,][]{Bullock:2000,Benson:2002} to the particle properties of dark matter \citep[\eg,][]{Ackermann:2015zua,Geringer-Sameth:2014qqa}.

%%% Observational review %%%

% 12 Classical Satellites (11 from Willman + CMa): LMC, SMC, For, Scl, Leo I, Leo II, UMi, Dra, Car, Sex, Sgr, (+ CMa)
% 15 Ultra-Faint Satellites (14 from Willman - Leo T + Psc II, Boo III): UMa, Wil 1, Boo I, UMa II, CVn, Seg 1, Com, Leo IV, CVn II, Her, (-Leo T), Boo II, Leo V, Seg 2, (+ Boo III, + Psc II)
% Willman notes that Wil 1, Seg 2, and Boo II have not been confirmed, but Josh feels otherwise...
Milky Way satellite galaxies are typically discovered as arcminute-scale statistical overdensities of individually resolved stars in wide-field optical imaging surveys \citep[][and references therein]{Willman:2010}.
Prior to 2005, there were twelve known ``classical'' Milky Way satellite galaxies with absolute magnitudes in the range $-8 \magn \gtrsim M_V \gtrsim -18 \magn$ \citep{McConnachie:2012a}.
From 2005 to 2014, \CHECK{fifteen} ultra-faint satellite galaxies with $M_V \gtrsim -8 \magn$ were identified in Sloan Digital Sky Survey \citep[SDSS;][]{York:2000} data through a combination of systematic searches
\citep{2005AJ....129.2692W,2005ApJ...626L..85W,2006ApJ...650L..41Z,2006ApJ...643L.103Z,2006ApJ...647L.111B,2007ApJ...654..897B,2008ApJ...686L..83B,2009MNRAS.397.1748B,2010ApJ...712L.103B,2006ApJ...645L..37G,2009ApJ...693.1118G,2006ApJ...653L..29S,2007ApJ...656L..13I,2007ApJ...662L..83W}
and dedicated spectroscopic follow-up observations 
\citep{2005ApJ...630L.141K,2006ApJ...650L..51M,2007MNRAS.380..281M,2007ApJ...670..313S,2009ApJ...692.1464G,2009ApJ...690..453K,2009MNRAS.397.1748B,2009ApJ...694L.144W,2009ApJ...702L...9C,2009A&A...506.1147A,2011AJ....142..128W,2011ApJ...736..146K,2011ApJ...733...46S,2013ApJ...770...16K}.
Several outer halo star clusters and/or more compact stellar systems of uncertain classification were reported during roughly the same period \citep[\eg,][]{2007ApJ...669..337K,2009AJ....137..450W,2010ApJ...712L.103B,2012ApJ...753L..15M,2013ApJ...767..101B,Laevens:2014a, 2014MNRAS.441.2124B,2015ApJ...799...73K,2015ApJ...803...63K}.
Dwarf galaxies are distinguished from star clusters by having a dynamical mass that is substantially larger than the mass inferred from the luminous stellar population and/or a significant dispersion in stellar metallicities indicative of multiple generations of star formation and a gravitational potential deep enough to retain supernova ejecta \citep{2012AJ....144...76W}.

% 14 new stellar systems: 8 from Bechtol - (Kim 2) + (Gru I, Tri II, Hyd II, Peg III, Hor II, Dra II, Sgr II)
Since the beginning of 2015, a combination of new optical imaging surveys and a reanalysis of SDSS data resulted in the detection of \ndsphs additional Milky Way companions, tentatively classified as dwarf galaxies \citep{Bechtol:2015wya, Koposov:2015cua,  Laevens:2015a, martin_2015_hydra_ii, kim_2015_pegasus_iii, Kim:2015c, Laevens:2015b}.
The new galaxy candidates have been examined for their (lack of) neutral gas \citep{Westmeier:2015}, possible association with the Magellanic Clouds \citep{Deason:2015b,Yozin:2015}, spatial distribution within the Galactic halo \citep{Pawlowski:2015dua}, and have become targets for indirect dark matter searches \citep{Drlica-Wagner:2015xua,Geringer-Sameth:2015lua,Hooper:2015ula}.
The provisional classification of these stellar systems as dwarf galaxies relies on their low surface brightnesses, large physical sizes, large ellipticities, and/or large heliocentric distances.
However, some of these objects lie in a region of size-luminosity space where the distinction between dwarf galaxy and globular cluster is ambiguous.
Thus far, spectroscopic observations have confirmed that three of the recently discovered satellites, Reticulum~II \citep{Simon:2015fdw, walker_2015_ret_ii,Koposov:2015b}, Horologium~I \citep{Koposov:2015b}, and Hydra~II \citep{Kirby:2015a} possess the kinematic and chemical signatures of galaxies.

\begin{figure*}
  \includegraphics[width=1.\textwidth]{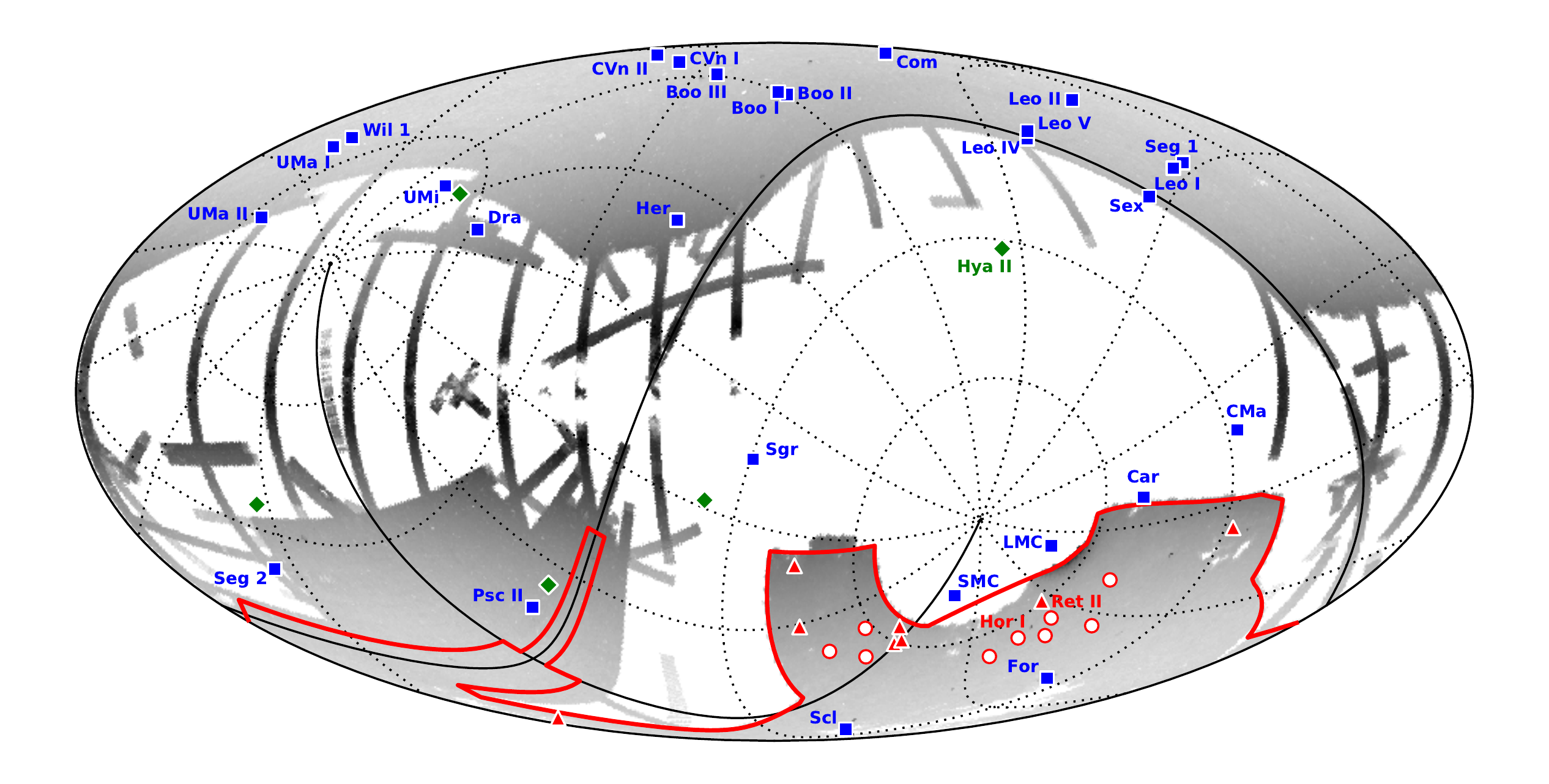}
  \caption{
Locations of the \ntotal new dwarf galaxy candidates reported here (red triangles) along with \nyearI previously reported dwarf galaxy candidates in the DES footprint \citep[red circles;][]{Bechtol:2015wya, Koposov:2015cua, Kim:2015c}, five recently discovered dwarf galaxy candidates located outside the DES footprint \citep[green diamonds;][]{Laevens:2015a, martin_2015_hydra_ii, kim_2015_pegasus_iii,Laevens:2015b}, and twenty-seven Milky Way satellite galaxies known prior to 2015 \citep[blue squares;][]{McConnachie:2012a}.
  Systems that have been confirmed as satellite galaxies are individually labeled.
  The figure is shown in Galactic coordinates (Mollweide projection) with the coordinate grid marking the equatorial coordinate system (solid lines for the equator and zero meridian). 
  The gray scale indicates the logarithmic density of stars with $r < 22$ from SDSS and DES.
  The two-year coverage of DES is $\roughly 5000\deg^2$ and nearly fills the planned DES footprint (outlined in red). 
  For comparison, the Pan-STARRS 1 $3\pi$ survey covers the region of sky with $\delta_{2000} > -30^{\circ}$ \citep{Laevens:2015b}.
}
  \label{fig:skymap}
\end{figure*}

The Dark Energy Survey \citep[DES;][]{Abbott:2005bi}, which began science operations in late 2013, has already had a large impact on our census of Milky Way substructures. 
In the first year of DES data, \nyearI new dwarf galaxy candidates were discovered in a region spanning less than 10\% of the southern hemisphere \citep{Bechtol:2015wya,Koposov:2015cua,Kim:2015c}.%
\footnote{The system Kim~2/\kimtwo/Indus~I was discovered independently in a separate data set by \citet{2015ApJ...803...63K} slightly before the DES announcement.  
With deeper observations, \citeauthor{2015ApJ...803...63K} concluded that this object is likely a star cluster, and we do not include it in our count of dwarf galaxy candidates.}
Here, we present an extension of the analysis described in \citet{Bechtol:2015wya} incorporating the second year of DES observations to expand the survey coverage from $\roughly 1800 \deg^2$ to $\roughly 5000 \deg^2$ (\figref{skymap}).
In \secref{data}, we discuss the data reduction and catalog generation steps applied to the two-year DES data set and the resulting unique catalog of calibrated objects.
We review the various algorithms applied to the DES data to search for ultra-faint galaxies in \secref{methods} and describe our candidate selection criteria in \secref{selection}.
The \ntotal most significant stellar overdensities that are unassociated with previously known systems are reported in \secref{results}.
If these objects are confirmed to be ultra-faint dwarf galaxies, they will be named after their resident constellations --- Cetus~II (\cetII), Columba~I (\colI), Grus~II (\gruII), Indus~II (\indII),%
\footnote{To distinguish \indII from Kim~2/\kimtwo/Indus~I, we suggest that the new system be designated Indus~II if determined to be a dwarf galaxy.}
%\footnote{As mentioned before, the designation of \kimtwo/Indus~I by \citet{Bechtol:2015wya} and \citet{Koposov:2015cua} has been superseded by the earlier discovery by \citet{2015ApJ...803...63K}.}
 Reticulum~III (\retIII), Tucana~III (\tucIII), Tucana~IV (\tucIV), and Tucana~V (\tucV) --- and named DES 2 through $N$ if found to be globular clusters \cite[e.g.,][]{Luque:2015}.
We place the DES candidates in context with other known Local Group galaxies and the Magellanic system in \secref{discussion}, and we conclude in \secref{conclusions}.

\section{Data Set}
\label{sec:data}

DES is an ongoing optical imaging survey of $\roughly 5000 \deg^2$ in the south Galactic cap using the Dark Energy Camera \citep[DECam;][]{flaugher_2015_decam} on the 4-m Blanco Telescope at Cerro Tololo Inter-American Observatory (CTIO).
The DECam focal plane comprises 62 2k$\times$4k CCDs dedicated to science imaging and 12 2k$\times$2k CCDs for guiding, focus, and alignment. 
The DECam field-of-view covers $3\deg^2$ with a central pixel scale of $0\farcs263$.
DES is scheduled for 525 nights distributed over five years, during which period each point in the survey footprint will be imaged ten times in each of the $grizY$ bands \citep{Abbott:2005bi}.

The searches presented in \citet{Bechtol:2015wya} and \citet{Luque:2015} utilized an object catalog generated from the coadded images in the DES year-one annual release (Y1A1).
To expedite the search for Milky Way satellites in year two, the present analysis uses data products derived from single-epoch imaging instead. 
This data set is referred to as the DES year-two quick release (Y2Q1), and its construction is summarized below.

\subsection{DES Year-Two Quick Release}
\label{sec:y2q1}

{\it Observations:} The Y2Q1 data set consists of \CHECK{26,590} DECam exposures taken during the first two years of DES observing that pass DES survey quality cuts. 
Slightly less than half of the DES survey area was observed during the first season (Y1; 2013 August 15 to 2014 February 9) with typically two to four overlapping exposures, referred to as ``tilings,'' in each of the $grizY$ filters.
The second season (Y2; 2014 August 7 to 2015 February 15) covered much of the remaining survey area to a similar depth.
Exposures taken in the $griz$ bands are 90s, while $Y$-band exposures are 45s.

{\it Image Reduction:} The DES exposures were processed with the DES data management (DESDM) image detrending pipeline \citep[][Gruendl et al., in preparation]{Sevilla:2011,2012SPIE.8451E..0DM,2012ApJ...757...83D}. 
This pipeline corrects for cross-talk between CCD amplifier electronics, bias level variations, and pixel-to-pixel sensitivity variations (flat-fielding). 
Additional corrections are made for nonlinearity, fringing, pupil, and illumination. 
Both the Y1 and Y2 exposures were reduced with the same image detrending pipeline used to process Y1A1.

{\it Single-Epoch Catalog Generation}: Astronomical source detection and photometry were performed on a per exposure basis using the \code{PSFex} and \code{SExtractor} routines \citep{2011ASPC..442..435B, 1996A&AS..117..393B} in an iterative process \citep[][Gruendl et al., in preparation]{2012SPIE.8451E..0DM}. 
As part of this step, astrometric calibration was performed with \code{SCAMP} \citep{Bertin:2006} by matching objects to the UCAC-4 catalog \citep{Zacharias:2013}. 
The \code{SExtractor} source detection threshold was set at $S/N > 10$ for the Y1 exposures, while for Y2 this threshold was lowered to $S/N > 5$ (the impact of this change on the stellar completeness is discussed in \secref{star_selection}).
During the catalog generation process, we flagged problematic images (\eg, CCDs suffering from reflected light, imaging artifacts, point-spread function (PSF) mis-estimation, \etc) and excluded the affected objects from subsequent analyses.
The resulting photometric catalogs were ingested into a high-performance relational database system.
Throughout the rest of this paper, photometric fluxes and magnitudes refer to \SExtractor output for the PSF model fit.

{\it Photometric Calibration:} Photometric calibration was performed using the stellar locus regression technique \citep[SLR:][]{Ivezic:2004a, MacDonald:2004a,High:2009a,Gilbank:2011a,2012ApJ...757...83D,Coupon:2012a,Kelly:2014a}.
Our reference stellar locus was empirically derived from the globally-calibrated DES Y1A1 stellar objects in the region of the Y1A1 footprint with the smallest $E(B-V)$ value from the \citet[SFD;][]{1998ApJ...500..525S} interstellar extinction map.
We performed a $1\arcsec$ match on all Y1 and Y2 objects with $S/N > 10$ observed in $r$-band and at least one other band.
We then applied a high-purity stellar selection based on the weighted average of the \spreadmodel quantity for the matched objects ($|\wavgspreadmodel[r]| < 0.003$; see below).
The average zero point measured in Y1A1, ${\rm ZP}_{grizY} = \{30.0,30.3,30.2,29.9,28.0\}$, was assigned to each star as an initial estimate.
Starting from this coarse calibration, we began an iterative procedure to fix the color uniformity across the survey footprint.
We segmented the sky into equal-area pixels using the \HEALPix scheme~\citep{2005ApJ...622..759G}. 
For each $\roughly 0.2 \deg^2$ (resolution $nside=128$) \HEALPix pixel, we chose the DES exposure in each band with the largest coverage and ran a modified version of the \code{Big MACS} SLR code \citep{Kelly:2014a}\footnote{\url{https://code.google.com/p/big-macs-calibrate/}} to calibrate each star from the reference exposure with respect to the empirical stellar locus.
These stars became our initial calibration standards.
We then adjusted the zero points of other CCDs so that the magnitudes of the matched detections agreed with the calibration set from the reference exposure.
CCDs with fewer than 10 matched stars or with a large dispersion in the magnitude offsets of matched stars ($\sigma_{\rm ZP} > 0.1 \magn$) were flagged.
For each calibration star, we computed the weighted-average magnitude in each band using these new CCD zero points; this weighted-average magnitude was used as the calibration standard for the next iteration of the SLR. 
In the first iteration, we assigned SLR zero points to the calibration stars based on the \HEALPix pixel within which they reside. 
In subsequent iterations, we assigned SLR zero points to the calibration stars based on a bi-linear interpolation of their positions onto the \HEALPix grid of SLR zero points. 
After the second iteration, the color zero points were stable at the $1-2 \mmag$ level.
The absolute calibration was set against the 2MASS $J$-band magnitude of matched stars (making use of the stellar locus in color-space), which were de-reddened using the SFD map with a reddening law of $A_J = 0.709 \times E(B-V)_{\rm SFD}$ from \citet{Schlafly:2011}.
The resulting calibrated DES magnitudes are thus already corrected for Galactic reddening by the SLR calibration.

{\it Unique Catalog Generation:}
There is significant imaging overlap within the DES footprint. 
For our final high-level object catalog, we identified unique objects by performing a $0\farcs5$ spatial match across all five bands.
For each unique object, we calculated the magnitude in each filter in two ways: (1) taking the photometry from the single-epoch detection in the exposure with the largest effective depth,%
\footnote{The effective depth is calculated from the exposure time, sky brightness, seeing, and atmospheric transmission \citep{Neilsen:2015}.}
and (2) calculating the weighted average ($wavg$) magnitude from multiple matched detections.
During the catalog generation stage, we kept all objects detected in any two of the five filters (this selection is later restricted to the $g$- and $r$-bands for our dwarf galaxy search).
The process of combining catalog-level photometry increases the photometric precision, but does not increase the detection depth.

The resulting Y2Q1 catalog covers $\roughly 5100 \deg^2$ in any single band, $\roughly 5000 \deg^2$ in both $g$- and $r$-band, and $\roughly 4700\deg^2$ in all five bands.
The coverage of the Y2Q1 data set is shown in \figref{skymap}.
The Y2Q1 catalog possesses a median relative astrometric precision of $\roughly 40 \mas$ per coordinate and a median absolute astrometric uncertainty of $\roughly 140 \mas$ per coordinate when compared against UCAC-4.
The median absolute photometric calibration agrees within $\roughly3\%$ with the de-reddened global calibration solution of Y1A1 in the overlap region in the $griz$ bands. 
When variations in the reddening law are allowed, this agreement improves to $\lesssim 1\%$.
The color uniformity of the catalog is $\roughly 1\%$ over the survey footprint (tested independently with the red sequence of galaxies).
The Y2Q1 catalog has a median point-source depth at a $S/N = 10$ of $g=23.4$, $r=23.2$, $i=22.4$, $z=22.1$, and $Y=20.7$.

\begin{figure*}[t]
  \includegraphics[width=0.5\textwidth]{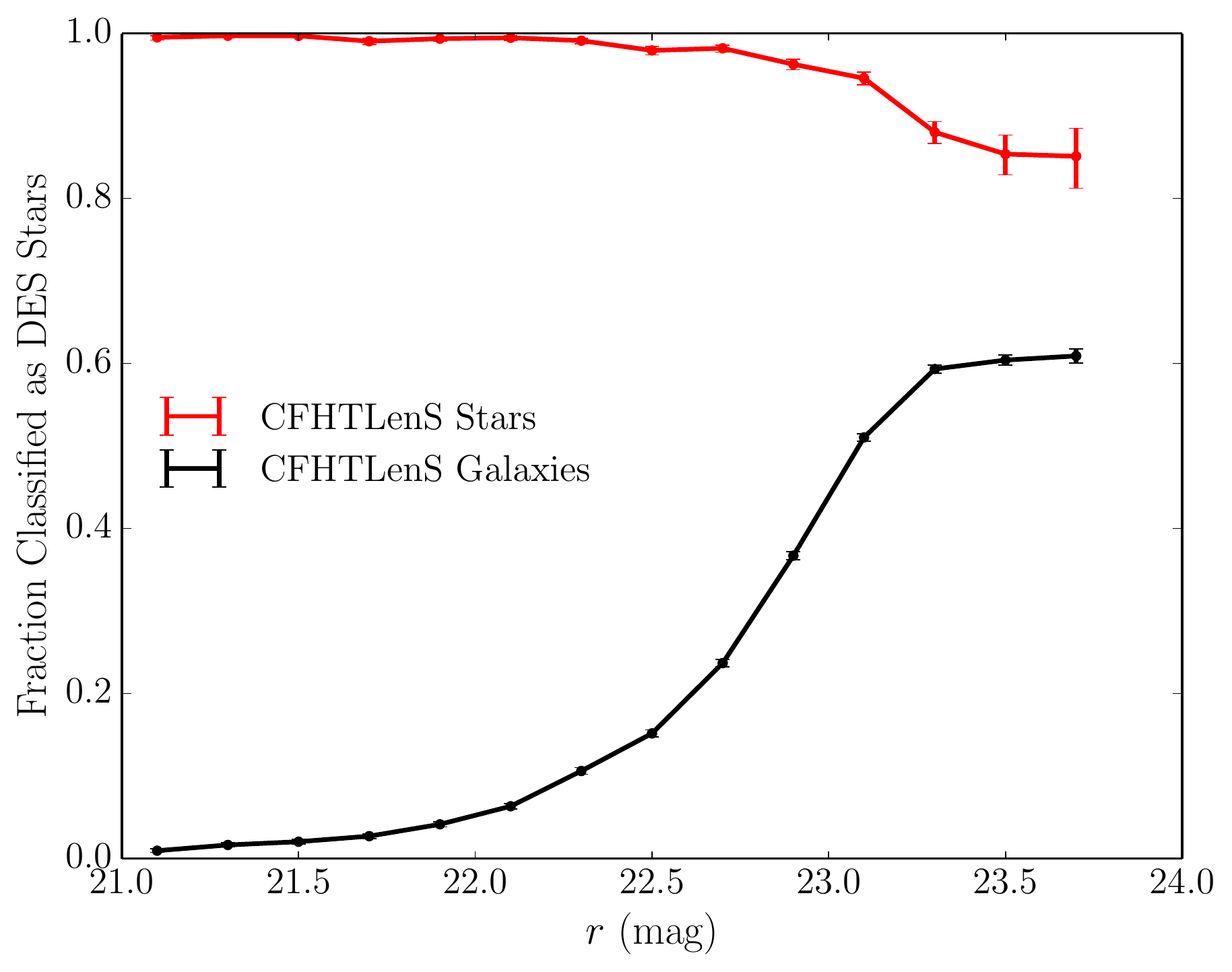}
  \includegraphics[width=0.5\textwidth]{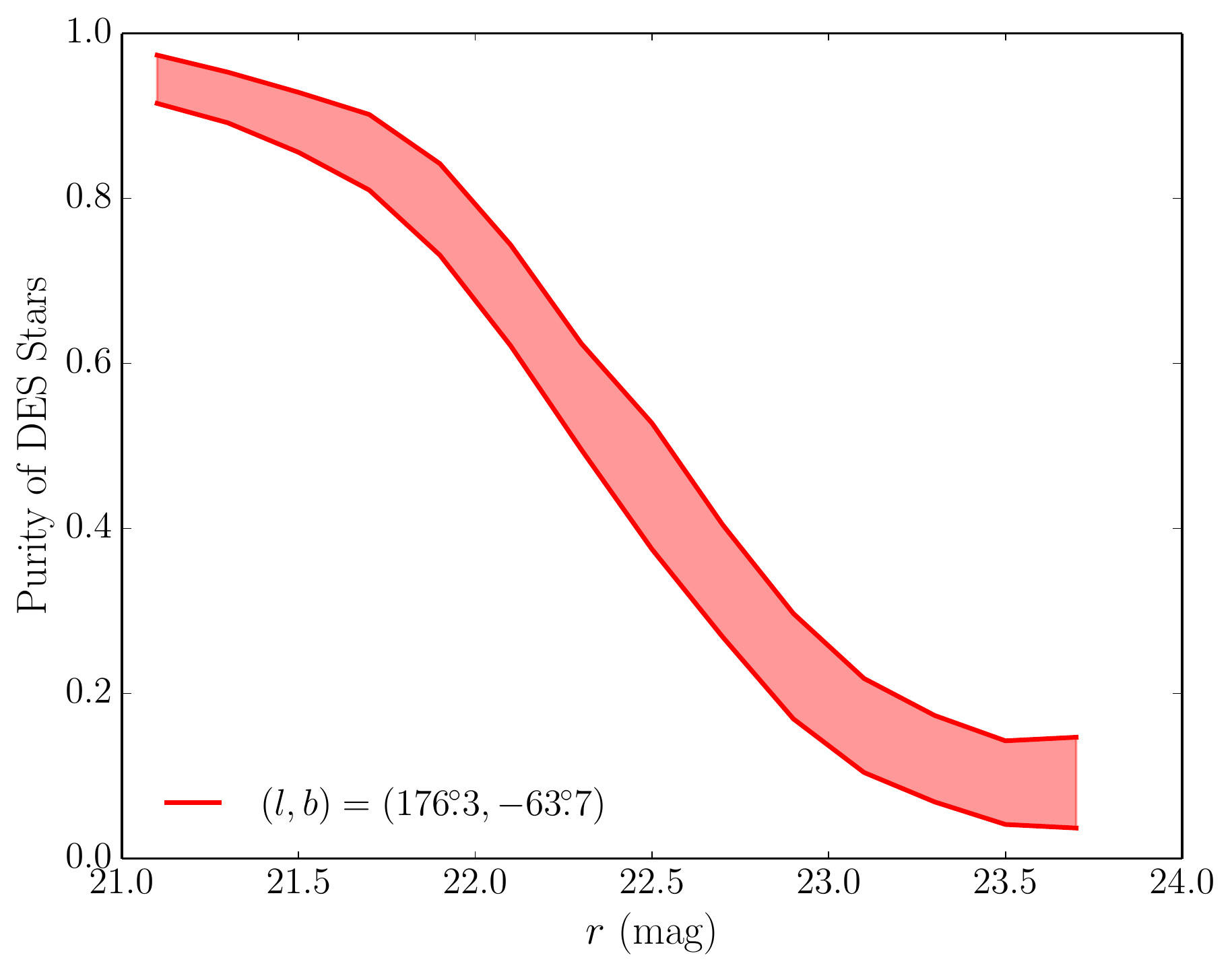}
  \caption{Performance of the morphological stellar classification based on $|\wavgspreadmodel[r]| \allowbreak < \allowbreak 0.003 + \spreaderrmodel[r]$ was evaluated by comparing against deeper CFHTLenS imaging (\secref{star_selection}). \textit{Left}: Fraction of CFHTLenS-classified stars and galaxies that pass our DES stellar selection criteria. \textit{Right}: Stellar purity evaluated in a typical high-Galactic latitude field centered at Galactic coordinates $(l, b) = (176\fdg3, -63\fdg7)$. The shaded region represents the uncertainty associated with objects that are not confidently classified as either stars or galaxies with CFHTLenS. If a large fraction of these ambiguous objects are true stars, the inferred stellar purity for DES is near the upper bound of the indicated range.
}
  \label{fig:stellar}
\end{figure*}

\subsection{Stellar Object Sample} 
\label{sec:star_selection}

We selected stars observed in both the $g$- and $r$-bands from the Y2Q1 unique object catalog following the common DES prescription based on the 
\wavgspreadmodel and \spreaderrmodel morphological quantities~\citep[\eg,][]{Chang:2015,Koposov:2015cua}.
Specifically, our stellar sample consists of well-measured objects with $|\wavgspreadmodel[r]| \allowbreak < \allowbreak 0.003 + \spreaderrmodel[r]$, $\flags[\{g,r\}] < 4$, and $\magerrpsf[\{g,r\}] < 1$. 
By incorporating the uncertainty on \spreadmodel, it is possible to maintain high stellar completeness at faint magnitudes while achieving high stellar purity at bright magnitudes. 
We sacrificed some performance in our stellar selection by using the $r$-band \spreadmodel which has a slightly worse PSF than the $i$-band; however, relaxing the requirement on $i$-band coverage increased the accessible survey area.

The stellar completeness is the product of the object detection efficiency and the stellar classification efficiency.
As described in \secref{y2q1}, different source detection thresholds were applied to the individual Y1 and Y2 exposures, and accordingly, the detection efficiency for faint objects in Y2Q1 varies across the survey footprint.
We used a region of the DES Science Verification data set (SVA1), imaged to full survey depth (\ie, $\roughly 10$ overlapping exposures in each of $grizY$), to estimate the object detection completeness using the individual Y1 exposures.
For the Y2 exposures, we instead compared to publicly available catalogs from the Canada-France-Hawaii Telescope Lensing Survey \citep[CFHTLenS;][]{2013MNRAS.433.2545E}.
In both cases, the imaging for the reference catalog is $\roughly 2 \magn$ deeper than the Y2Q1 catalog.
These comparisons show that the Y2Q1 object catalog is $> 90\%$ complete at about $0.5 \magn$ brighter than the $S/N = 10$ detection limit.

We estimated the stellar classification efficiency by comparing to CFHTLenS and by using statistical estimates from the Y2Q1 data alone.
For the CFHTLenS matching, we identified probable stars in the CFHTLenS catalog using a combination of the \classstar output from \SExtractor and the \var{fitclass} output from the \code{lensfit} code \citep{2012MNRAS.427..146H}.
Specifically, our CFHTLenS stellar selection was $(\var{fitclass} = 1)\,{\rm OR}\,(\classstar > 0.98)$, and our galaxy selection was $(\var{fitclass} = 0)\,{\rm OR}\,(\classstar < 0.2)$.
Comparing against the DES stellar classification based on \spreadmodel, we found that the Y2Q1 stellar classification efficiency exceeds 90\% for $r < 23 \magn$, and falls to $\roughly 80$\% by $r = 24$ (\figref{stellar}).
We independently estimated the stellar classification efficiency by creating a test sample of Y2Q1 objects with high stellar purity using the color-based selection $r - i > 1.7$.
We then applied the \spreadmodel morphological stellar selection to evaluate the stellar completeness for the test sample.
The stellar selection efficiency determined by the color-based selection test is in good agreement with that from the CFHTLenS comparison.

For a typical field at high Galactic latitude, the stellar purity is $\roughly 50$\% at $r = 22.5 \magn$, and falls to $\roughly 10$\% at the faint magnitude limit (\figref{stellar}).
In other words, the majority of faint objects in the ``stellar sample'' are misclassified galaxies rather than foreground halo stars.\footnote{This conclusion differs from that of \citet{Koposov:2015cua}, who state that low levels of contamination are observed to the magnitude limit of DES.}
This transition can be seen in the color-magnitude distribution diagnostic plots for individual candidates (\secref{snowflake}).

\section{Search Methods}
\label{sec:methods}

In this section we describe three complementary search strategies used to identify overdensities of individually resolved stars matching the old and metal-poor populations characteristic of ultra-faint galaxies.
Each technique used a different method to estimate the spatially varying field density and to select potential satellite member stars.
The output of each search method was a set of seed coordinates, (\ra, \dec, \modulus), corresponding to statistical excesses relative to the local field density.
These seeds were cross-matched between methods and examined in greater detail (\secref{results}).

When searching for new stellar systems, each detection algorithm applied a deep magnitude threshold ($g < 23.5 \magn$ or $g < 24 \magn$ depending on the search technique).
At these faint magnitudes, our stellar sample is far from complete (\secref{star_selection}).
However, real stellar systems are dominated by faint stars, and the signal-to-noise ratio for satellite discovery can increase by including objects that are fainter than the stellar completeness limit.
After candidate systems were identified, we applied a brighter magnitude threshold ($g < 23 \magn$) to avoid biasing the fit results during object characterization (\secref{results}).

One potential concern when extending our search to very faint magnitudes is the increased contamination from misclassified galaxies (\secref{star_selection}).
However, as long as the contaminating objects have a similar statistical distribution in the target region and the surrounding area, each technique naturally incorporates the increased object density into the background estimate.
There are situations where the assumption of uniformity does not apply (\eg, clusters of galaxies, variations in survey depth, gaps in coverage, \etc); therefore, when assembling our candidate list, we also examined the distribution of probable galaxies as well as stars with colors far from the expected isochrone (\secref{results}).

We briefly describe our three search strategies below.

\subsection{Stellar Density Maps}
\label{sec:stellar_density_maps}

Our most straightforward and model-independent search technique used a simple isochrone filter to facilitate visual inspection of the stellar density field.
The specific algorithm was a variant of the general approach described in Section 3.1 of \citet{Bechtol:2015wya} and follows from the methods of \citet{2008ApJ...686..279K}, \citet{2009AJ....137..450W}, and \citet{2015ApJ...799...73K}.

We first increased the contrast of putative satellite galaxies by selecting stars that were consistent with the isochrone of an old ($\age = 12 \Gyr$) and metal-poor ($\metal = 0.0001$) stellar population \citep{Dotter:2008}.
Specifically, we selected stars within $0.1\magn$ of the isochrone locus in color-magnitude space, enlarged by the photometric measurement uncertainty of each star added in quadrature.
After isochrone filtering, we smoothed the stellar density field with a $2 \arcmin$ Gaussian kernel.
For each density peak in the smoothed map, we scanned over a range of radii ($1\arcmin$ to $18 \arcmin$) and computed the Poisson significance of finding the observed number of stars within each circle centered on the density peak given the local field density.
This process was repeated for stellar populations at a range of distance moduli ($16 \magn \leq \modulus \leq 24 \magn$) and diagnostic plots were automatically generated for each unique density peak.
Our nominal magnitude threshold for this search was $g < 24 \magn$; however, in regions on the periphery of the survey with large variations in coverage we used a brighter threshold of $g < 23 \magn$.

\subsection{Matched-filter Search}
\label{sec:sparsex}
Our second search strategy utilized a matched-filter algorithm \citep{Luque:2015,Balbinot:2011,2013ApJ...767..101B}.
We began by binning objects in our stellar sample with $17 \magn < g < 24 \magn$  into $1 \arcmin \times 1 \arcmin$ spatial pixels and $g$ versus $g-r$ color-magnitude bins of $0.01 \magn \times 0.05 \magn$.
We then created a grid of possible simple stellar populations modeled with the isochrones of \citet{Bressan:2012} and populated according to a \citet{2001ApJ...554.1274C} initial mass function to use as signal templates.
The model grid spanned a range of distances ($10\kpc < D < 200 \kpc$), ages ($9 < \log_{10} \age < 10.2$), and metallicities ($Z = \{0.0002,0.001, 0.007\}$).
The color-magnitude distribution of the field population was empirically determined over larger $10\degree \times 10\degree$ regions.
For each isochrone in the grid, we then fit the normalization of the simple stellar population in each spatial pixel to create a corresponding density map for the full Y2Q1 area.

The set of pixelized density maps do not assume a spatial model for the stellar system, and can be used to search for a variety of stellar substructures (\eg, streams and diffuse clouds).
To search specifically for dwarf galaxies, we convolved the density maps with Gaussian kernels having widths ranging from $0 \arcmin$ (no convolution) to $9 \arcmin$ and used \SExtractor to identify compact stellar structures in those maps.
This convolution and search was performed on each map for our grid of isochrones.
The resulting seeds were ranked according to their statistical significance and by the number of maps in which they appeared, i.e., the number of isochrones for which an excess was observed.

\subsection{Likelihood-based Search}
\label{sec:likelihood}

Our third search technique was a maximum likelihood-based algorithm that simultaneously used the full spatial and color-magnitude distribution of stars, as well as photometric uncertainties and information about the survey coverage.
The likelihood was constructed from the product of Poisson probabilities for detecting each individual star given its measured properties and the parameters that describe the field population and putative satellite \citep[Section 3.2 of][]{Bechtol:2015wya}.
When searching for new stellar systems, we used a spherically symmetric Plummer model as the spatial kernel, and a composite isochrone model consisting of four isochrones of different ages, $\age = \{ 12 \Gyr, 13.5 \Gyr \}$, and metallicities, $\metal = \{ 0.0001,0.0002 \}$, to bracket a range of possible stellar populations \citep{Bressan:2012}.
We scanned the Y2Q1 data testing for the presence of a satellite galaxy at each location on a multi-dimensional grid of sky position (0\farcm7 resolution; \HEALPix $nside = 4096$), distance modulus ($16 \magn <\modulus< 23 \magn$), and kernel half-light radius ($r_h = \{0\fdg03, 0\fdg1, 0\fdg3 \}$).
For the likelihood search, we restricted our magnitude range to $16 \magn < g < 23.5 \magn$ as a compromise between survey depth and stellar completeness.
The statistical significance of a putative galaxy at each grid point was expressed as a test statistic (\TS) based on the likelihood ratio between a hypothesis that includes a satellite galaxy versus a field-only hypothesis~\citep[Equation 4 of][]{Bechtol:2015wya}.
As a part of the model fitting, the satellite membership probability is computed for every star in the region of interest.
These photometric membership probabilities can be used to visualize which stars contribute to the candidate detection significance (\secref{results}) and have previously been used to select targets for spectroscopic follow-up~\citep{Simon:2015fdw,walker_2015_ret_ii}.

\newcommand{\detcaption}{Detection of ultra-faint galaxy candidates.\label{tab:detection}}
\newcommand{\detcomments}{
Characteristics of satellite galaxy candidates discovered in DES Y2 data.
Best-fit parameters from the maximum-likelihood analysis assume a \citet{Bressan:2012} isochrone.
 Uncertainties come from the highest density interval containing \interval of the posterior distribution.
The uncertainty on the distance modulus (\modulus) also includes systematic uncertainties coming from the choice of theoretical isochrone and photometric calibration (\secref{results}).
The azimuthally averaged half-light radius ($r_h$) is quoted for all candidates.
For systems with evidence for asphericity (Bayes' factor $> 3$), we quote the ellipticity ($\epsilon$), the position angle ($\phi$), and the length of the semi-major axis of the ellipse containing half of the light ($a_h = r_h / \sqrt{1 - \epsilon}$). 
Upper limits on the ellipticity are quoted for other candidates at \limit confidence. 
``Map Sig.'' refers to detection significance of the candidate from the stellar density map search method (\secref{stellar_density_maps}). 
``TS Scan'' refers to the significance \citep[Equation 4 in][]{Bechtol:2015wya} from the likelihood scan (\secref{likelihood}). 
$\Sigma p_i$ is the estimated number of satellite member stars in the DES Y2Q1 catalog with $g < 23 \magn$.
}
\newcommand{\tablenotes}{
\tablenotetext{a}{Fit with a spherically symmetric Plummer profile due to the possible  presence of tidal tails (\secref{snowflake}).}
\tablenotetext{b}{Fit with a composite isochrone: $\age = \{ 12 \Gyr, 13.5 \Gyr \}$, $\metal = \{ 0.0001, 0.0002 \}$ (\secref{results}).}

}

\begin{\tabletype}{l ccccccccccc }
%\tablecolumns{13}
%\tablewidth{0pt}
\tabletypesize{\tiny}
\tablecaption{ \detcaption }
\tablehead{
Name & $\alpha_{2000}$ & $\delta_{2000}$ & $m-M$ & $r_{h}$ & $a_{h}$ & $\epsilon$ & $\phi$ & Map Sig. & TS Scan & $\sum p_i$ \\ 
 & (deg) & (deg) &  & $(\arcmin)$ & $(\arcmin)$ &  &  & $(\sigma)$ &  & 
}
\startdata
\gruII (Gru\,II)   & 331.02 & -46.44 & $18.62 \pm 0.21$ & $6.0^{+0.9}_{-0.5}$ & \ldots & $<0.2$ & \ldots & 15.7 &  369 &  161 \\
\tucIII (Tuc\,III)\tablenotemark{a} & 359.15 & -59.60 & $17.01 \pm 0.16$ & $6.0^{+0.8}_{-0.6}$ & \ldots & \ldots & \ldots & 11.1 &  390 &  168 \\
\colI (Col\,I)\tablenotemark{b} &  82.86 & -28.03 & $21.30 \pm 0.22$ & $1.9^{+0.5}_{-0.4}$ & \ldots & $<0.2$ & \ldots & 10.5 &   71 &   33 \\
\tucIV (Tuc\,IV)   &   0.73 & -60.85 & $18.41 \pm 0.19$ & $9.1^{+1.7}_{-1.4}$ & $11.8^{+2.2}_{-1.8}$ & $0.4^{+0.1}_{-0.1}$ & $ 11 \pm 9$ &  8.7 &  287 &  134 \\
\retIII (Ret\,III)\tablenotemark{b} &  56.36 & -60.45 & $19.81 \pm 0.31$ & $2.4^{+0.9}_{-0.8}$ & \ldots & $<0.4$ & \ldots &  8.1 &   56 &   22 \\
\tucV (Tuc\,V)     & 354.35 & -63.27 & $18.71 \pm 0.34$ & $1.0^{+0.3}_{-0.3}$ & $1.8^{+0.5}_{-0.6}$ & $0.7^{+0.1}_{-0.2}$ & $ 30 \pm 5$ &  8.0 &  129 &   24 \\
\vspace{-0.2cm}\\\tableline\tableline\vspace{-0.2cm}\\
\indII (Ind\,II)\tablenotemark{b} & 309.72 & -46.16 & $21.65 \pm 0.16$ & $2.9^{+1.1}_{-1.0}$ & \ldots & $<0.4$ & \ldots &  6.0 &   32 &   22 \\
\cetII (Cet\,II)   &  19.47 & -17.42 & $17.38 \pm 0.19$ & $1.9^{+1.0}_{-0.5}$ & \ldots & $<0.4$ & \ldots &  5.5 &   53 &   21 \\

\enddata
{\footnotesize \tablecomments{ \detcomments }}
\tablenotes
\end{\tabletype}

\section{Candidate Selection}
\label{sec:selection}

The search methods described in \secref{methods} each produced a set of seed locations (\ra, \dec, \modulus) for overdensities in the stellar field and a significance associated with each seed.
Before performing a computationally-intensive multi-parameter likelihood fit to characterize each seed, we applied a set of simple selection cuts to remove seeds that are unlikely to be new Milky Way satellites. 
We set detection significance thresholds at ${>}\,5.5 \sigma$ for the stellar density map search and $\TS > 45$ ($\roughly 6\sigma$) for the maximum-likelihood search. 
For the matched-filter method, the ten highest-ranked seeds were selected from each $10 \degree \times 10 \degree$ search region.
After thresholding, the union of all three search techniques yielded several hundred unique seeds.
Many of these seeds were attributed to steep gradients in the stellar density field, numerical effects near the survey boundaries, and imaging artifacts.
We also compared against catalogs of other astrophysical objects that produce false positives, such as large nearby galaxies~\citep{1973ugcg.book.....N,2004yCat.7239....0H} and galaxy clusters \citep{2014ApJ...785..104R}.
These objects were removed from our list of candidates and we did not pursue investigation at lower significance thresholds due to an increased false positive rate.

\begin{figure*}
  \includegraphics[width=1.\textwidth]{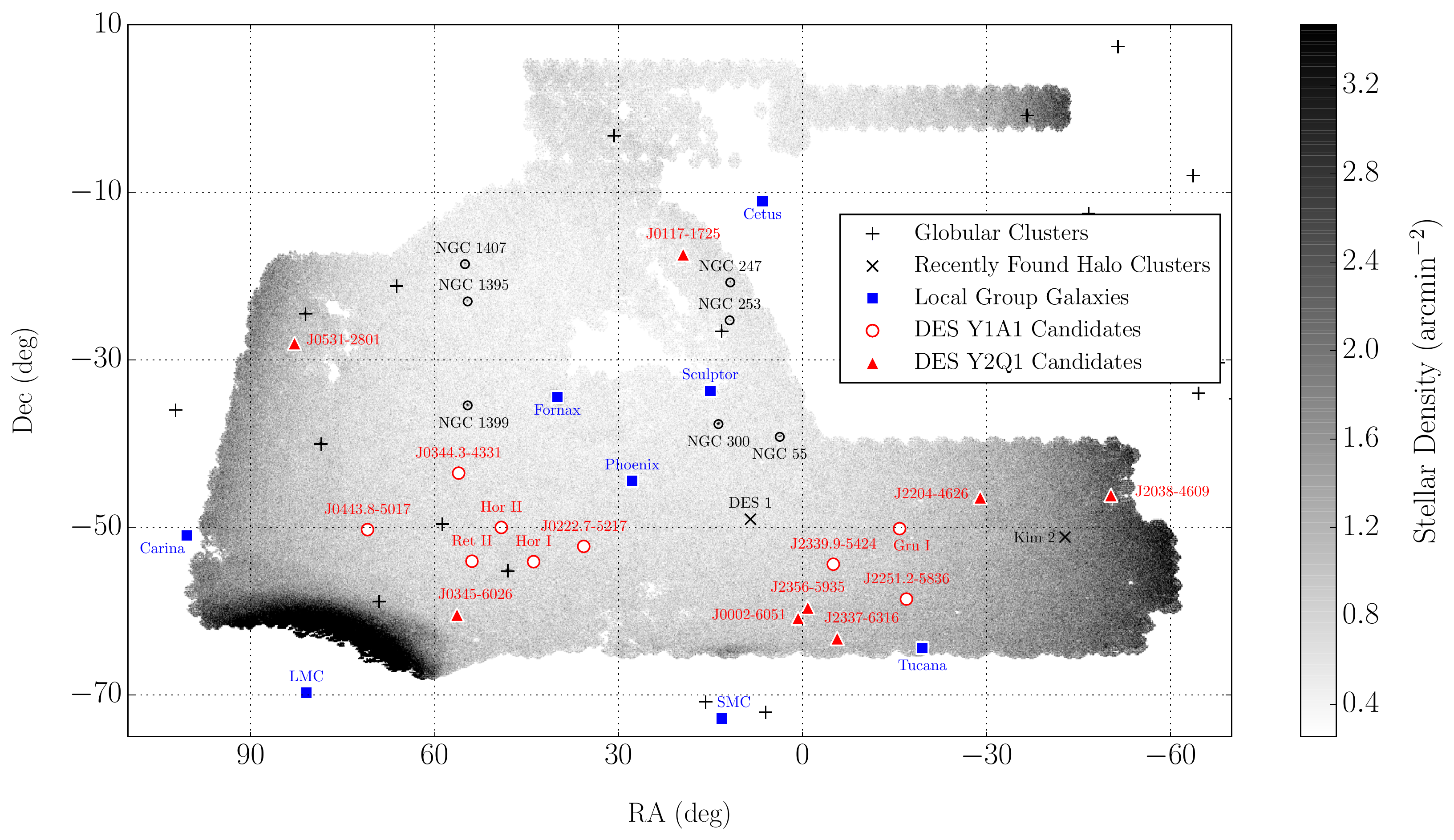}
  \caption{Cartesian projection of the density of stars observed in both $g$- and $r$-bands with $g < 23$ and $g - r < 1$ over the DES Y2Q1 footprint ($\roughly 5000\deg^2$). 
Globular clusters are marked with ``$+$'' symbols \citep[][2010 edition]{Harris96}, two faint outer halo clusters are marked with ``$\times$'' symbols \citep{2015ApJ...803...63K,Luque:2015}, Local Group galaxies known prior to DES are marked with blue squares \citep{McConnachie:2012a}, dwarf galaxy candidates discovered in Y1 DES data are marked with red outlined circles, while the Y2 stellar systems are marked with red triangles. 
The periphery of the LMC can be seen in the southeast corner of the footprint, while the Galactic stellar disk can be seen on the eastern and western edges.}
\label{fig:skymap_zoom}
\end{figure*}

The resulting list of seeds was matched against catalogs of known Local Group galaxies \citep{McConnachie:2012a} and star clusters \citep[][2010 edition]{2013A&A...558A..53K,Harris96}. 
Some of the most prominent stellar systems in our list of seeds were the Fornax, Phoenix, Sculptor,\footnote{The center of Sculptor lies in a hole in Y2Q1 coverage, though the outskirts are still detected at high significance.} and Tucana dwarf galaxies, and the globular clusters AM\,1, Eridanus, Reticulum, Whiting 1, \ngc{288}, \ngc{1261}, \ngc{1851}, \ngc{1904}, and \ngc{7089}.
Additionally, seed locations were compared against other stellar overdensities recently reported in the DES footprint \citep[][]{2015ApJ...803...63K,Bechtol:2015wya,Koposov:2015cua,Kim:2015c,Luque:2015} and all \nsys objects were recovered with high significance.
The locations of previously known stellar systems in the DES footprint are shown in \figref{skymap_zoom}.

As an additional test of the physical nature of the $\roughly30$ highest significance unassociated seeds, we explored the posterior probability density of the likelihood with an affine invariant Markov Chain Monte Carlo (MCMC) ensemble sampler \citep[MCMC; ][]{2013PASP..125..306F}.\footnote{\emcee v2.1.0: \url{http://dan.iel.fm/emcee/}}
This step was meant as a preliminary investigation of whether the posterior distribution was well constrained by the data.
We fixed the distance modulus of each object at the value that maximized the likelihood in the grid search (\secref{likelihood}) and sampled the morphological parameters (center position, extension, ellipticity, and position angle) along with the richness of the putative stellar system.
We imposed a Jeffreys' prior on the extension ($\mathcal{P}(a_h) \propto 1/a_h$) and flat priors on all other parameters.
The posterior probability was sampled with $2.5\times10^5$ steps and we rejected seeds that lacked a well-defined maximum in the posterior.

The process described above resulted in the selection of \ntotal candidate stellar systems.
\Nobjs of these candidates, \gruII (Gru~I), \tucIII (Tuc~III), \colI (Col~I), \tucIV (Tuc~IV), \retIII (Ret~III), and \tucV (Tuc~V), are high-confidence detections in clean regions of the survey.
The additional \ncand candidates, \indII (Ind~II) and \cetII (Cet~II), are lower confidence and reside in more complicated regions.
In  \tabref{detection} we report the coordinates and detection significances of each of these objects.

\section{Candidate Characterization}
\label{sec:results}

In this section, we describe the iterative procedure used to characterize each of the \ntotal candidate stellar systems.
When fitting the new candidates, we applied a brighter magnitude threshold, $g < 23 \magn$, to mitigate the impact of stellar incompleteness.
The results of our characterization are shown in \tabref{detection} and \tabref{properties}.

We began by simultaneously fitting the morphological parameters and distance modulus of each candidate following the procedure described in \secref{selection}.
Best-fit values were derived from the marginalized posterior distribution and the morphological parameters were temporarily fixed at these values. 
We then ran a MCMC chain simultaneously sampling the distance, age, metallicity, and richness, assuming flat priors on each parameter.
Significant correlations between these parameters were found, and in some cases the age and metallicity were poorly constrained (see below).
To assess the error contribution from intrinsic uncertainty in the isochrone, we resample the posterior distribution using the Dartmouth isochrone family \citep{Dotter:2008}.
In general, the best-fit distance moduli agree to within 0.1 \magn when the data were fit by these two isochrone classes.
To determine the uncertainty on the distance modulus quoted in \tabref{detection} we started with the highest posterior density interval from the \citet{Bressan:2012} isochrone fit. 
We then calculated a systematic component from the difference in the best-fit distance modulus derived with \citet{Bressan:2012} and \citet{Dotter:2008} isochrones. 
An additional $\pm 0.03 \magn$ uncertainty was added to account for uncertainty in the photometric calibration.
The age and metallicity values were taken from the peak of the marginalized posterior density from the \citet{Bressan:2012} isochrone fit.

For the distant systems (\ie, \colI, \indII, and \retIII), the main-sequence turnoff (MSTO) is fainter than the $g < 23 \magn$ limit imposed for our fit.
In these systems, the data provide weaker constraints on the age and metallicity.
We therefore fit the distance modulus using a composite of four \citet{Bressan:2012} isochrones with fixed ages, $\age = \{ 12 \Gyr, 13.5 \Gyr \}$, and metallicities, $\metal = \{ 0.0001, 0.0002 \}$.
We followed the procedure above to incorporate systematic uncertainties from the choice of theoretical isochrone family and from the calibration.\footnote{One notable exception is the case of \indII, where significant constraining power comes from a set of horizontal branch (HB) stars. 
For \indII we estimated the distance uncertainty from the \citet{Bressan:2012} isochrone analysis alone.}

Fixing the distance, age, and metallicity at the values derived in the previous step, we then re-fit the morphological parameters.
The best-fit values of the morphological parameters and their highest posterior density intervals are listed in \tabref{detection}.
The majority of objects show no significant evidence for ellipticity, which is confirmed by calculating the Bayes factor (BF) via the Savage-Dickey density ratio~\citep{Dickey:1971,Trotta:2005ar}.
For objects with evidence for asphericity ($BF > 3$), we report both the elliptical Plummer radius, $a_h$, and the position angle, $\phi$.
For all objects, we report the azimuthally averaged half-light radius, $r_h$.
In the region of \tucIII there is a linear structure visible in the filtered stellar density map, consistent with a set of tidal tails (\secref{snowflake}). 
We require the spatial profile of \tucIII to be azimuthally symmetric in the MCMC fit and analyze this linear feature separately.

The best-fit parameters derived from our iterative MCMC analysis along with several additional derived properties are reported in \tabref{detection} and \tabref{properties}.
The physical size of each object was calculated by propagating the uncertainty from the distance and azimuthally averaged half-light radius.
Absolute magnitudes were calculated according to the prescription of \citet{2008ApJ...684.1075M} and do not include the distance uncertainty.

The two lower confidence candidates are located in complicated regions of the survey. 
\cetII is located in a region of sparse coverage and is close to both a CCD chip gap and a larger hole in the survey. 
\indII is located in a region of non-uniform depth at the interface of Y1 and Y2 observations.
Although we have attempted to account for these issues in the likelihood fit, we caution that the parameters derived for these systems may be less secure than for the other candidates.

As a final note, the iterative MCMC fitting procedure described above was also applied to the satellite galaxy candidate Grus~I \citep{Koposov:2015cua}.
Grus~I was identified in Y1A1 imaging data, but was located close to a CCD chip gap in a region with sparse coverage.
We reanalyzed Grus~I with the additional Y2 exposures and found that its structural parameters are consistent within uncertainties with those reported by \citet{Koposov:2015cua}.

\newcommand{\propcaption}{Derived properties of ultra-faint galaxy candidates.\label{tab:properties}}
\newcommand{\propcomments}{
Derived properties of the DES Y2 satellite candidates.
Stellar masses (\Mstar) are computed for a \citet{2001ApJ...554.1274C} initial mass function.
The absolute visual magnitude is derived via the procedure of \citet{2008ApJ...684.1075M} using the transformation equations from \citet{Bechtol:2015wya}.
The uncertainty on the azimuthally-averaged physical half-light radius ($r_{1/2}$) includes the uncertainty in the projected half-light radius and distance.
Age (\age) and metallicity (\metal) values come from the peak of the posterior distribution.
}

\begin{\tabletype}{l ccccccccc }
\tablecolumns{13}
\tablewidth{0pt}
\tabletypesize{\tiny}
\tablecaption{ \propcaption }
\tablehead{
Name & $\ell$ & $b$ & Distance & \Mstar & $M_{V}$ & $r_{1/2}$ & $\tau$ & $Z$ \\ 
 & (deg) & (deg) & (kpc) & ($10^3 \Msun$) & (mag) & (pc) & (Gyr) & 
}
\startdata
\gruII (Gru\,II)   & 351.15 & -51.94 & $  53 \pm 5$ & $ 3.4^{+0.3}_{-0.4}$ & $-3.9 \pm 0.22$ & $  93 \pm 14$ & $12.5$ & $0.0002$ \\
\tucIII (Tuc\,III)\tablenotemark{a} & 315.38 & -56.19 & $  25 \pm 2$ & $ 0.8^{+0.1}_{-0.1}$ & $-2.4 \pm 0.42$ & $  44 \pm 6$ & $10.9$ & $0.0001$ \\
\colI (Col\,I)\tablenotemark{b} & 231.62 & -28.88 & $ 182 \pm 18$ & $ 6.2^{+1.9}_{-1.0}$ & $-4.5 \pm 0.17$ & $ 103 \pm 25$ & \ldots & \ldots \\
\tucIV (Tuc\,IV)   & 313.29 & -55.29 & $  48 \pm 4$ & $ 2.2^{+0.4}_{-0.3}$ & $-3.5 \pm 0.28$ & $ 127 \pm 24$ & $11.6$ & $0.0001$ \\
\retIII (Ret\,III)\tablenotemark{b} & 273.88 & -45.65 & $  92 \pm 13$ & $ 2.0^{+0.6}_{-0.7}$ & $-3.3 \pm 0.29$ & $  64 \pm 24$ & \ldots & \ldots \\
\tucV (Tuc\,V)     & 316.31 & -51.89 & $  55 \pm 9$ & $ 0.5^{+0.1}_{-0.1}$ & $-1.6 \pm 0.49$ & $  17 \pm 6$ & $10.9$ & $0.0003$ \\
\vspace{-0.2cm}\\\tableline\tableline\vspace{-0.2cm}\\
\indII (Ind\,II)\tablenotemark{b} & 353.99 & -37.40 & $ 214 \pm 16$ & $ 4.9^{+1.8}_{-1.6}$ & $-4.3 \pm 0.19$ & $ 181 \pm 67$ & \ldots & \ldots \\
\cetII (Cet\,II)   & 156.48 & -78.53 & $  30 \pm 3$ & $ 0.1^{+0.04}_{-0.04}$ & $ 0.0 \pm 0.68$ & $  17 \pm 7$ & $10.9$ & $0.0002$ \\

\enddata
{\footnotesize \tablecomments{ \propcomments }}
\tablenotes
\end{\tabletype}

\section{Discussion}
\label{sec:discussion}

\begin{figure}
  \includegraphics[width=\columnwidth]{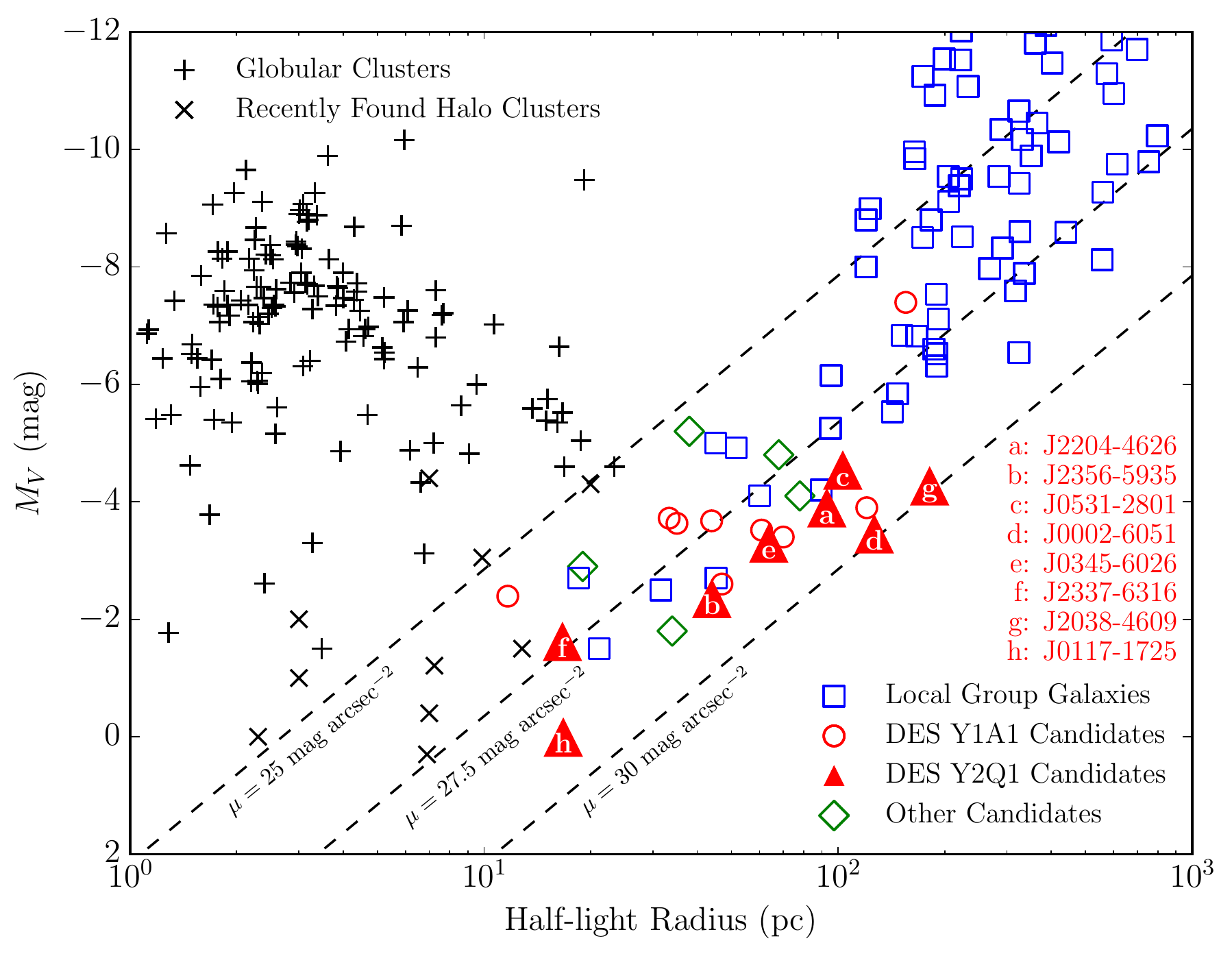}
  \caption{Local Group galaxies \citep{McConnachie:2012a} and globular clusters \citep[][2010 edition]{Harris96} occupy distinct regions in the plane of physical half-light radius (azimuthally averaged) and absolute magnitude. 
  The majority of DES satellite candidates (red triangles and circles) are more consistent with the locus of Local Group galaxies (blue squares) than with the population of Galactic globular clusters (black ``$+$'').
  Other recently reported dwarf galaxy candidates (green diamonds) include Hydra II \citep{martin_2015_hydra_ii}, Triangulum~II \citep{Laevens:2015a}, Pegasus~III \citep{kim_2015_pegasus_iii}, Draco~II and Sagittarius~II \citep{Laevens:2015b}.
  Several outer halo star clusters and systems of ambiguous classification are indicated with ``$\times$'' symbols: Koposov~1 and Koposov~2 \citep{2007ApJ...669..337K,2014AJ....148...19P}, Segue~3 \citep{2010ApJ...712L.103B,2011AJ....142...88F,2013MNRAS.433.1966O}, Mu\~{n}oz~1 \citep{2012ApJ...753L..15M}, Balbinot~1 \citep{2013ApJ...767..101B}, Laevens~1 \citep{Laevens:2014a,2014MNRAS.441.2124B}, Laevens~3 \citep{Laevens:2015b}, Kim~1 and Kim~2 \citep{2015ApJ...799...73K,2015ApJ...803...63K}, and \luque \citep{Luque:2015}.
  Dashed lines indicate contours of constant surface brightness at $\mu = \{25, 27.5, 30\} \magn \unit{arcsec}^{-2}$. 
}
\label{fig:size_luminosity}
\end{figure}

Spectroscopic observations are needed to definitively classify the newly discovered stellar systems as either star clusters or ultra-faint dwarf galaxies based upon their stellar dynamics and/or metallicity dispersions.
However, the distances, physical sizes, and luminosities derived from DES photometry (\tabref{properties}) already suggest a galactic classification for many of the new candidates.
In \figref{size_luminosity}, we show the distribution of physical half-light radius versus absolute magnitude for Milky Way globular clusters \citep[][2010 edition]{Harris96} and Local Group galaxies \citep{McConnachie:2012a}.
We find that all of the systems discovered in Y2Q1 fall along the established locus for nearby galaxies.
Several of the new systems possess lower surface brightnesses than any confirmed ultra-faint galaxy,%
\footnote{A potential exception to this statement is the large, faint, and perhaps tidally disrupted object B\"ootes~III \citep{2009ApJ...702L...9C}.}
supporting earlier conclusions that the threshold in surface brightness was an observational selection effect \citep{2008ApJ...686..279K,2009AJ....137..450W}.
Finally, globular clusters generally have small ellipticities, $\epsilon \lesssim 0.2$ \citep{2008AJ....135.1731V}, while dwarf galaxies are commonly more elliptical \citep{2008ApJ...684.1075M}.
However, the ellipticities of many of the new objects are not well constrained by the DES data and thus do not provide a strong indicator for the nature of these objects.
%Below we review some characteristics of the individual objects.

The spatial distribution of the new candidates within the DES footprint is suggestive of an association with the Magellanic Clouds (\secref{distribution}).
When discussing this scenario, the coordinates of the Large Magellanic Cloud (LMC) are taken to be $(\ra,\dec, D_{\odot}) = (80\fdg89,-69\fdg76,49.89\kpc)$ \citep[distance from][]{deGrijs:2014} and the coordinates of the Small Magellanic Cloud (SMC) are taken to be $(\ra,\dec,D_{\odot}) = (13\fdg19, -72\fdg83, 61.94\kpc)$ \citep[distance from][]{deGrijs:2015}.\footnote{Sky coordinates taken from NED: \url{https://ned.ipac.caltech.edu}.} 
While the new DES candidates reside in a region close to the Magellanic Stream, no candidate is found to be coincident with the main filament of the Stream.
A similar observation was made by \citet{Koposov:2015cua} for the DES stellar systems discovered in Y1.

We also investigate a potential group of satellites in the constellation Tucana, consisting of \tucIV and \tucV, and the object \tucII (Tuc~II) discovered in Y1 data \citep{Bechtol:2015wya,Koposov:2015cua}. 
The centroid of the Tucana group is at $(\ra,\dec,D_{\odot}) = (351\fdg90, -61\fdg06, 53.63\kpc)$, and the separation of each member from the centroid is $\lesssim 7 \kpc$ (\tabref{separations}).
This grouping of objects is projected onto a region of \ion{H}{i} gas that is likely a secondary filament or high-velocity cloud associated with the head of the Magellanic Stream \citep{Putman:2003,McClure-Griffiths:2009}.

\begin{figure}
  \includegraphics[width=\columnwidth]{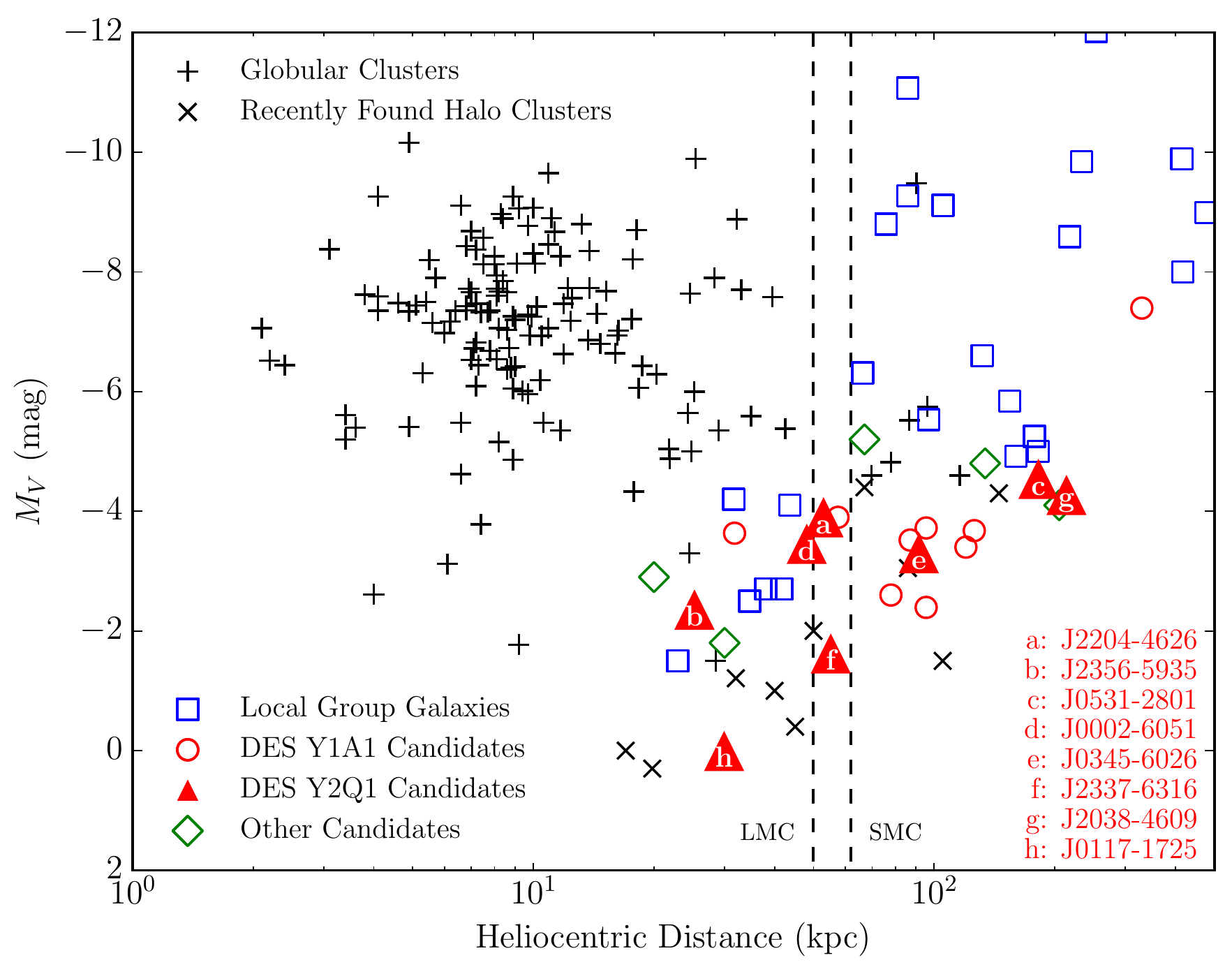}
  \caption{
DES satellite galaxy candidates (red circles and triangles) have comparable luminosities to known Local Group galaxies \citep[blue squares;][]{McConnachie:2012a} and other newly discovered galaxy candidates (green diamonds), but are typically found at larger distances.
%This suggests that DES may be probing the dwarf galaxy population over a larger volume than previous surveys.
Dashed lines indicate the heliocentric distances of the LMC and SMC.
Globular clusters \citep[black ``$+$'' symbols,][]{Harris96} and outer halo star clusters (``$\times$'' symbols) have been detected at comparable luminosities and distances due to their more compact nature.
The constituents of each source class can be found in \figref{size_luminosity}.
}
\label{fig:distance_luminosity}
\end{figure}

\subsection{Review of Individual Candidates}
\label{sec:snowflake}

\newcommand{\sepcaption}{Physical separations of ultra-faint galaxy candidates. \label{tab:separations}}
\newcommand{\sepcomments}{
Three-dimensional physical separation between DES satellite candidates and the LMC and SMC. 
Also listed is the heliocentric distance, the Galactocentric distance, and the separation to the centroid of the proposed Tucana group comprising \tucII, \tucIV, and \tucV.}

\begin{deluxetable}{l cccccc }
\tablecolumns{13}
\tablewidth{0pt}
\tabletypesize{\tiny}
\tablecaption{ \sepcaption }
\tablehead{
Name & $D_{\odot}$ & $D_{\rm GC}$ & $D_{\rm LMC}$ & $D_{\rm SMC}$ & $D_{\rm Tuc}$ \\ 
 & (kpc) & (kpc) & (kpc) & (kpc) & (kpc)
}
\startdata
\gruII (Gru\,II)   &  53 &  49 &  46 &  33 &  18 \\
\tucIII (Tuc\,III) &  25 &  23 &  32 &  38 &  28 \\
\colI (Col\,I)     & 182 & 186 & 149 & 157 & 168 \\
\tucIV (Tuc\,IV)   &  48 &  46 &  27 &  18 &   7 \\
\retIII (Ret\,III) &  92 &  92 &  45 &  40 &  53 \\
\tucV (Tuc\,V)     &  55 &  52 &  29 &  14 &   3 \\
\vspace{-0.2cm}\\\tableline\tableline\vspace{-0.2cm}\\
\indII (Ind\,II)    & 214 & 208 & 193 & 170 & 169 \\
\cetII (Cet\,II)   &  30 &  32 &  46 &  51 &  40 \\
\vspace{-0.2cm}\\\tableline\tableline\vspace{-0.2cm}\\
\tucII (Tuc\,II)   &  58 &  54 &  37 &  20 &   7 \\

\enddata
{\footnotesize \tablecomments{ \sepcomments }}

\end{deluxetable}

\begin{itemize}

\item {\bf \gruII} (Grus~II, \figref{cmd_gruII}): 
As the most significant new candidate, \gruII has \CHECK{$\roughly 161$} probable member stars with $g < 23 \magn$ in the DES imaging.
The large physical size of this system (\CHECK{$93\pc$}) allows it to be tentatively classified as an ultra-faint dwarf galaxy. 
A clear red giant branch (RGB) and MSTO are seen in the color-magnitude diagram of \gruII. 
Several likely blue horizontal branch (BHB) members are seen at $g \sim 19 \magn$.
There is a small gap in survey coverage $\roughly 0\fdg5$ from the centroid of \gruII, but this is accounted for in the maximum-likelihood analysis and should not significantly affect the characterization of this rich satellite.

\gruII is nearly equidistant from the LMC (\CHECK{$46\kpc$}) and from the Galactic center (\CHECK{49 \kpc}). 
It is slightly closer to the SMC (\CHECK{33\kpc}) and intriguingly close to the the group of stellar systems in Tucana (\CHECK{$\roughly 18\kpc$}).
While it is unlikely that \gruII is currently interacting with any of the other known satellites, we cannot preclude the possibility of a past association.

\begin{figure*}
  \includegraphics[width=1.\textwidth]{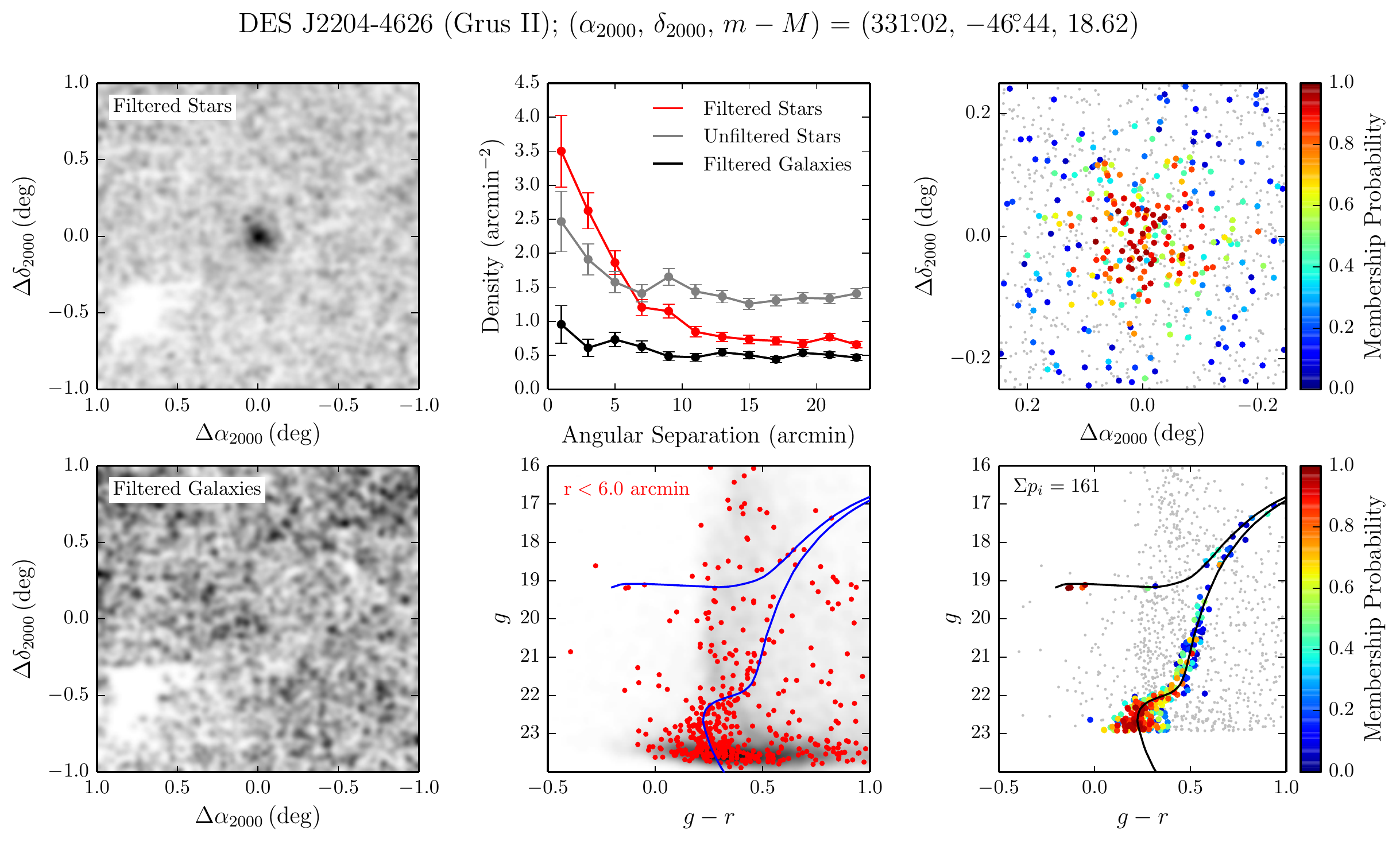}
  \caption{Stellar density and color-magnitude diagrams for \gruII (Grus~II). 
  \textit{Top left}: Spatial distribution of stellar objects with $g < 24 \magn$ that pass the isochrone filter (see text). The field of view is $2\degree \times 2\degree$ centered on the candidate and the stellar distribution has been smoothed by a Gaussian kernel with standard deviation $1\farcm2$.
  \textit{Top center}: Radial distribution of objects with $g - r < 1 \magn$ and $g < 24 \magn$: stars passing the isochrone filter (red), stars excluded from the isochrone filter (gray), and galaxies passing the isochrone filter (black).
   \textit{Top right}: Spatial distribution of stars with high membership probabilities within a $0\fdg5 \times 0\fdg5$ field of view. Small gray points indicate stars with membership probability less than 5\%.
   \textit{Bottom left}: Same as top left panel, but for probable galaxies that pass the isochrone filter.
   \textit{Bottom center}: The color-magnitude distribution of stars within $0\fdg1$ of the centroid are indicated with individual points. 
The density of the field within an annulus from radii of $0\fdg5$ to $1\fdg0$ is represented by the background two-dimensional histogram in grayscale. 
The blue curve shows the best-fit isochrone as described in \tabref{detection} and \tabref{properties}. 
   \textit{Bottom right}: Color-magnitude distribution of high membership probability stars.
  }
\label{fig:cmd_gruII}
\end{figure*}

\item {\bf \tucIII} (Tucana~III, \figref{cmd_tucIII}): 
\tucIII is another highly significant stellar system that presents a clearly defined main sequence and several RGB stars.
The physical size and luminosity of \tucIII (\CHECK{$r_{1/2} = 44\pc$}, \CHECK{$M_V = -2.4 \magn$}) are comparable to that of the recently discovered dwarf galaxy Reticulum~II. % \citep{Bechtol:2015wya,Koposov:2015cua}.
At a distance of \CHECK{$25\kpc$}, \tucIII would be among the nearest ultra-faint satellite galaxies known, along with Segue~1 (23\kpc), Reticulum~II (32\kpc), and Ursa Major~II (32\kpc).
The relative abundance of bright stars in \tucIII should allow for an accurate spectroscopic determination of both its velocity dispersion and metallicity. 
\tucIII is reasonably close to the LMC and SMC (32\kpc and 38\kpc, respectively), and measurements of its systemic velocity will help to elucidate a physical association.

As mentioned in \secref{results}, there is an additional linear feature in the stellar density field around \tucIII that strongly influences the fitted ellipticity of this galaxy candidate.
In \figref{tucIII_stream}, we show the $6 \degree \times 6\degree$ region centered on \tucIII which contains this linear feature.
Selecting stars with the same isochrone filter used to increase the contrast of \tucIII relative to the field population, we found a faint stellar overdensity extending $\roughly2\degree$ on both sides of \tucIII.
This feature has a FWHM of \CHECK{$\roughly0\fdg3$} (\ie, projected dimensions of \CHECK{$1.7\kpc \times 130\pc$} at a distance of \CHECK{$25\kpc$}) and is oriented \CHECK{$\roughly 85\degree$} east of north.
The length of the feature and the smooth density field observed with an inverted isochrone filter support the conclusion that this is a genuine stellar structure associated with \tucIII.

One interpretation is that \tucIII is in the process of being tidally stripped by the gravitational potential of the Milky Way. 
The relatively short projected length of the putative tails may imply that \tucIII is far from its orbital pericenter (where tidal effects are strongest) and/or that stripping began recently.
Despite the possible presence of tidal tails, the main body of \tucIII appears relatively round.
While the proximity of \tucIII  might make it an important object for indirect dark matter searches, evidence of tidal stripping would suggest that a large fraction of its outer dark matter halo has been removed.  
However, the mass within the stellar core (and therefore the dark matter content in the central region) can still likely be determined accurately \citep{Oh:1995, Munoz:2008, Penarrubia:2008}.

\begin{figure*}
  \includegraphics[width=\textwidth]{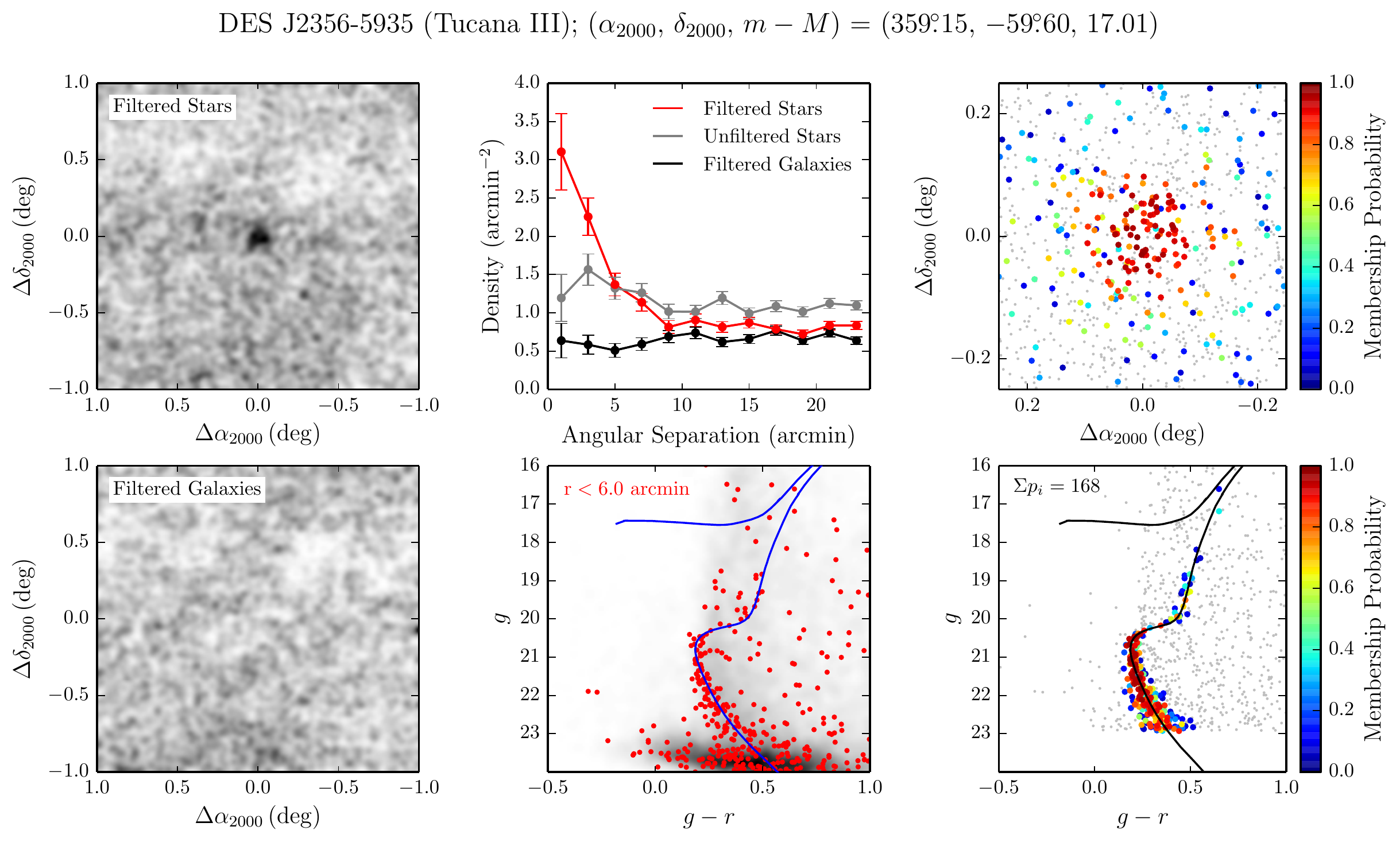}
  \caption{Analogous to \figref{cmd_gruII} but for \tucIII (Tucana~III).} 
\label{fig:cmd_tucIII}
\end{figure*}

\begin{figure*}
  \centering
  \includegraphics[width=0.7\textwidth]{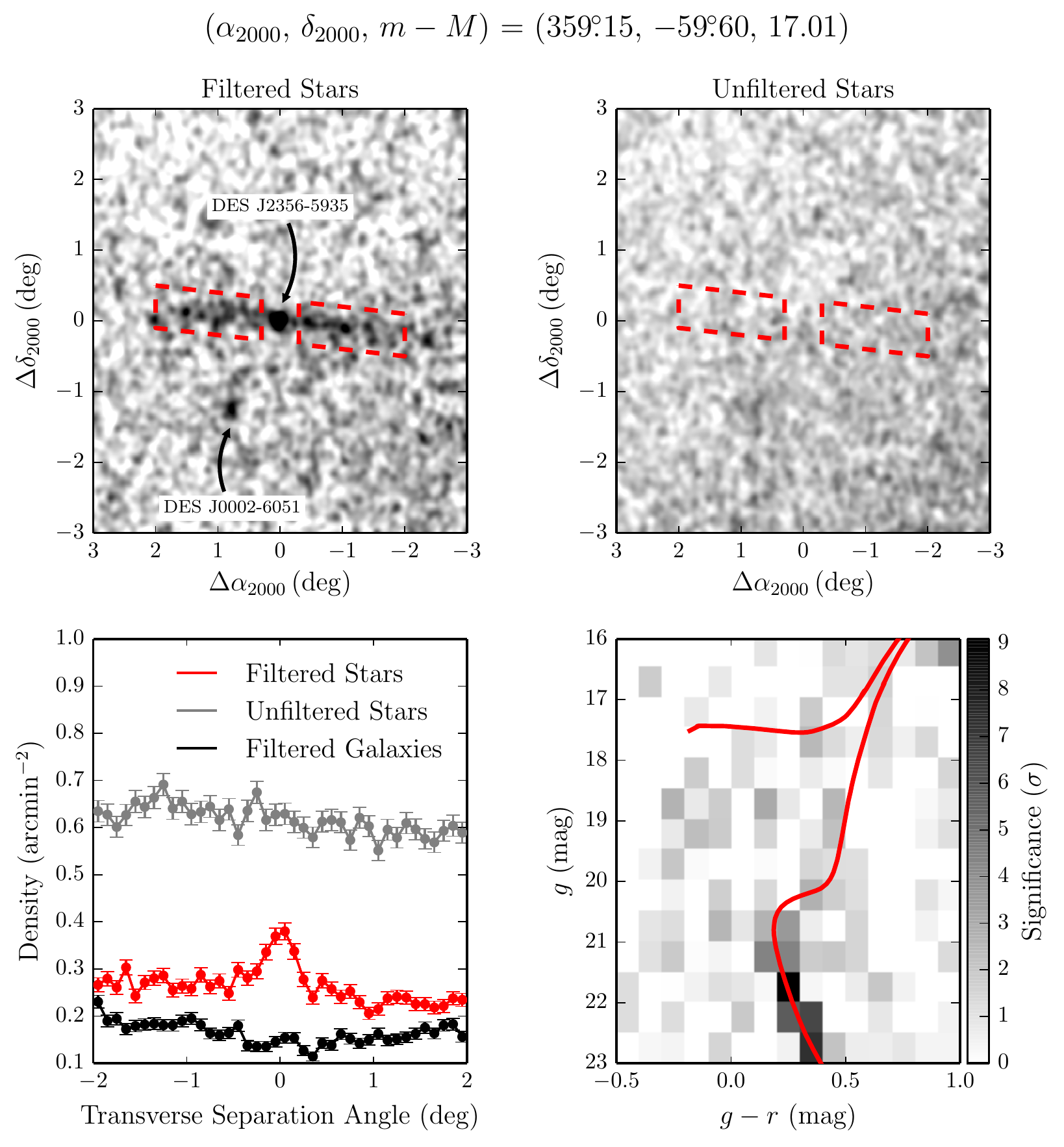}
  \caption{Possible tidal tails around \tucIII (Tucana~III). 
  \textit{Top left}: Spatial distribution of stellar objects with $g < 23 \magn$ that pass the same isochrone filter as applied to visualize \tucIII (\figref{cmd_tucIII}). Red boxes highlight the positions of the tidal tails.
  \textit{Top right}: Spatial distribution of stellar objects with $g - r < 1 \magn$ and $g < 23 \magn$ not passing the isochrone filter.
  \textit{Bottom left}: Average density profiles transverse to the tidal tail axis: stars passing the isochrone filter (red), other unfiltered stars (gray), and galaxies passing the isochrone filter (black).
  For the density calculation, the range of included angles parallel to the tidal tails is $\pm2\fdg0$, excluding the region within $0\fdg3$ of \tucIII.
  \textit{Bottom right}: Binned significance diagram representing the Poisson probability of detecting the observed number of stars within $0\fdg2$ of the tidal tail axis (excluding the region within $0\fdg3$ of \tucIII) for each bin of the color-magnitude space given the local field density.
  }
\label{fig:tucIII_stream}
\end{figure*}

\item {\bf \colI} (Columba~I, \figref{cmd_colI}): 
\colI is the second most distant of the Y2 DES candidates (\CHECK{182\kpc}) and is detected as a compact cluster of RGB stars.
Both the BHB and red horizontal branch (RHB) appear to be well populated.
The distance, physical size (\CHECK{103\pc}), and luminosity (\CHECK{$M_V = -4.5 \magn$}) of \colI place it in the locus of Local Group dwarf galaxies, comparable to Leo~IV and CVn~II \citep{2007ApJ...654..897B}.
\colI is isolated from the other new DES systems and is likely not associated with the Magellanic system.

\begin{figure*}
  \includegraphics[width=1.\textwidth]{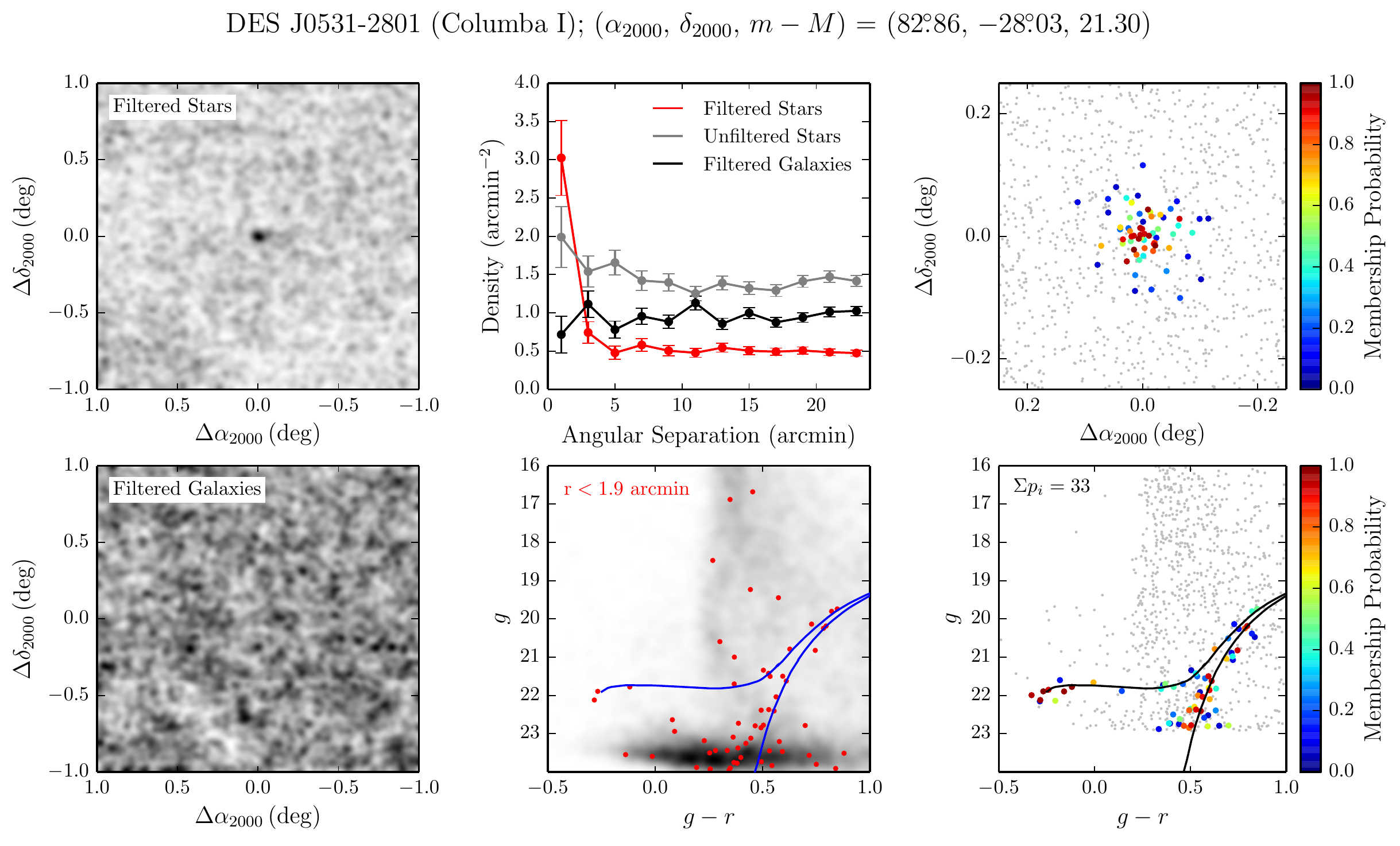}
  \caption{Analogous to \figref{cmd_gruII} but for \colI (Columba~I). 
  }
\label{fig:cmd_colI}
\end{figure*}

\item {\bf \tucIV} (Tucana~IV, \figref{cmd_tucIV}): 
\tucIV has the largest angular size of the candidates (\CHECK{$r_h = 9\farcm1$}), which at a distance of \CHECK{$48\kpc$} corresponds to a physical size of \CHECK{$r_{1/2} \sim 127 \pc$}. 
This large half-light radius is inconsistent with the sizes of known globular clusters \citep[][2010 edition]{Harris96}, thus making it very likely that \tucIV is a dwarf galaxy.
The measured ellipticity of \tucIV, \CHECK{$\epsilon = 0.4$}, is also consistent with a galactic classification.
\tucIV is found to be \CHECK{$27\kpc$} from the LMC and \CHECK{$18 \kpc$} from the SMC.
\tucIV is one of the proposed members of the Tucana group, with a centroid separation of \CHECK{$7 \kpc$}. 
Measurements of the radial velocity and proper motion of \tucIV will provide strong clues as to whether it was accreted as part of a system of satellites.
Similarly to \tucIII and \gruII, the MSTO of \tucIV is well-populated and clearly visible.
Several possible member stars can also be seen along the HB.

\begin{figure*}
  \includegraphics[width=1.\textwidth]{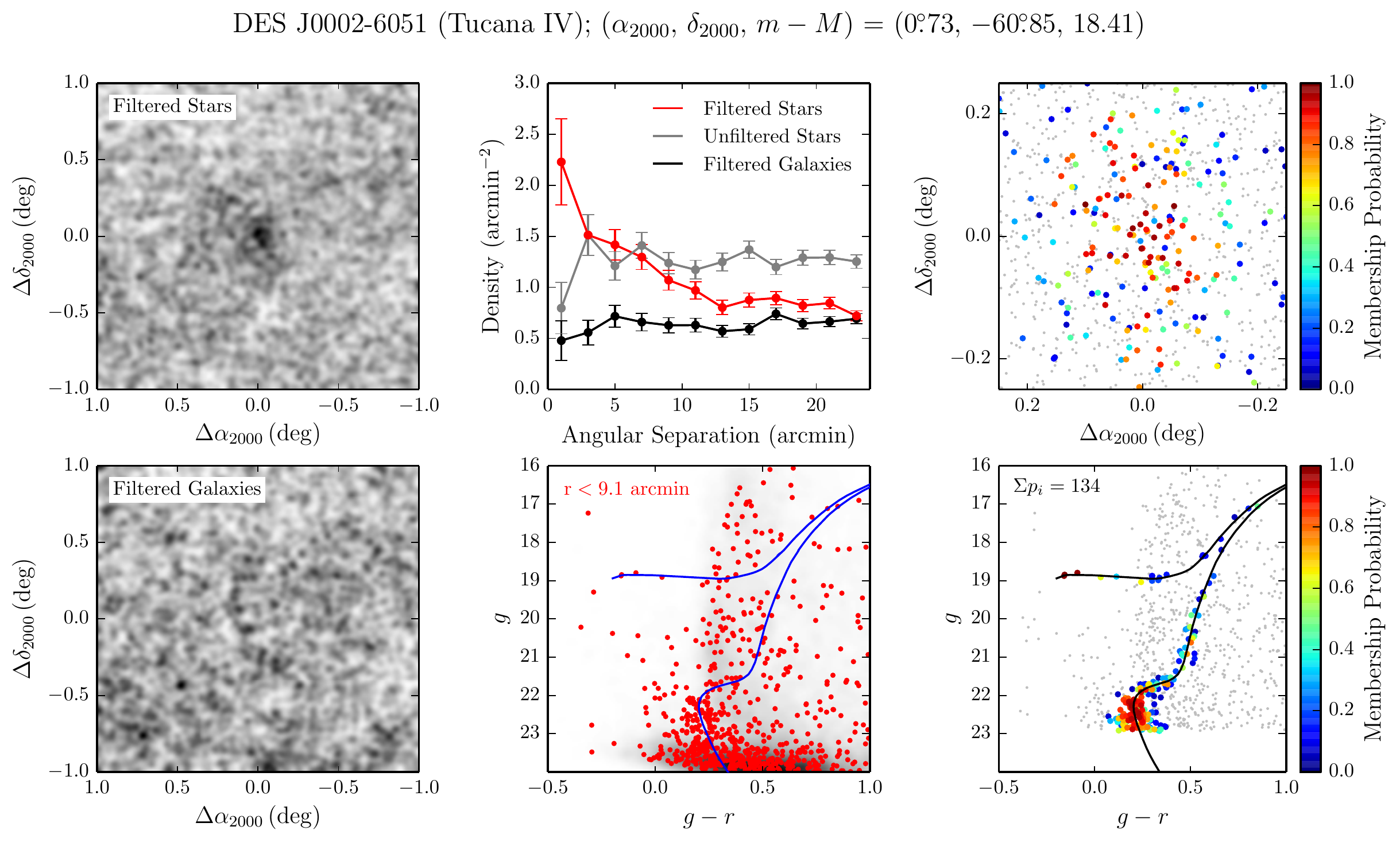}
  \caption{Analogous to \figref{cmd_gruII} but for \tucIV (Tucana~IV).
  }
\label{fig:cmd_tucIV}
\end{figure*}

\item {\bf \retIII} (Reticulum~III, \figref{cmd_retIII}): 
\retIII appears similar to \gruII in its structural properties, but is more distant (heliocentric distance of \CHECK{$92\kpc$}). 
Like \gruII, \retIII can be tentatively classified as a dwarf galaxy based on its physical size and low surface brightness.
There is some indication of asphericity for this object; however, the ellipticity is not significantly constrained by the DES data.
\retIII has a sparsely populated RGB, a few possible RHB members, and two possible BHB members.
The MSTO for \retIII is slightly fainter than our fitting threshold of $g < 23 \magn$, thus its age is poorly constrained.

\begin{figure*}
  \includegraphics[width=1.\textwidth]{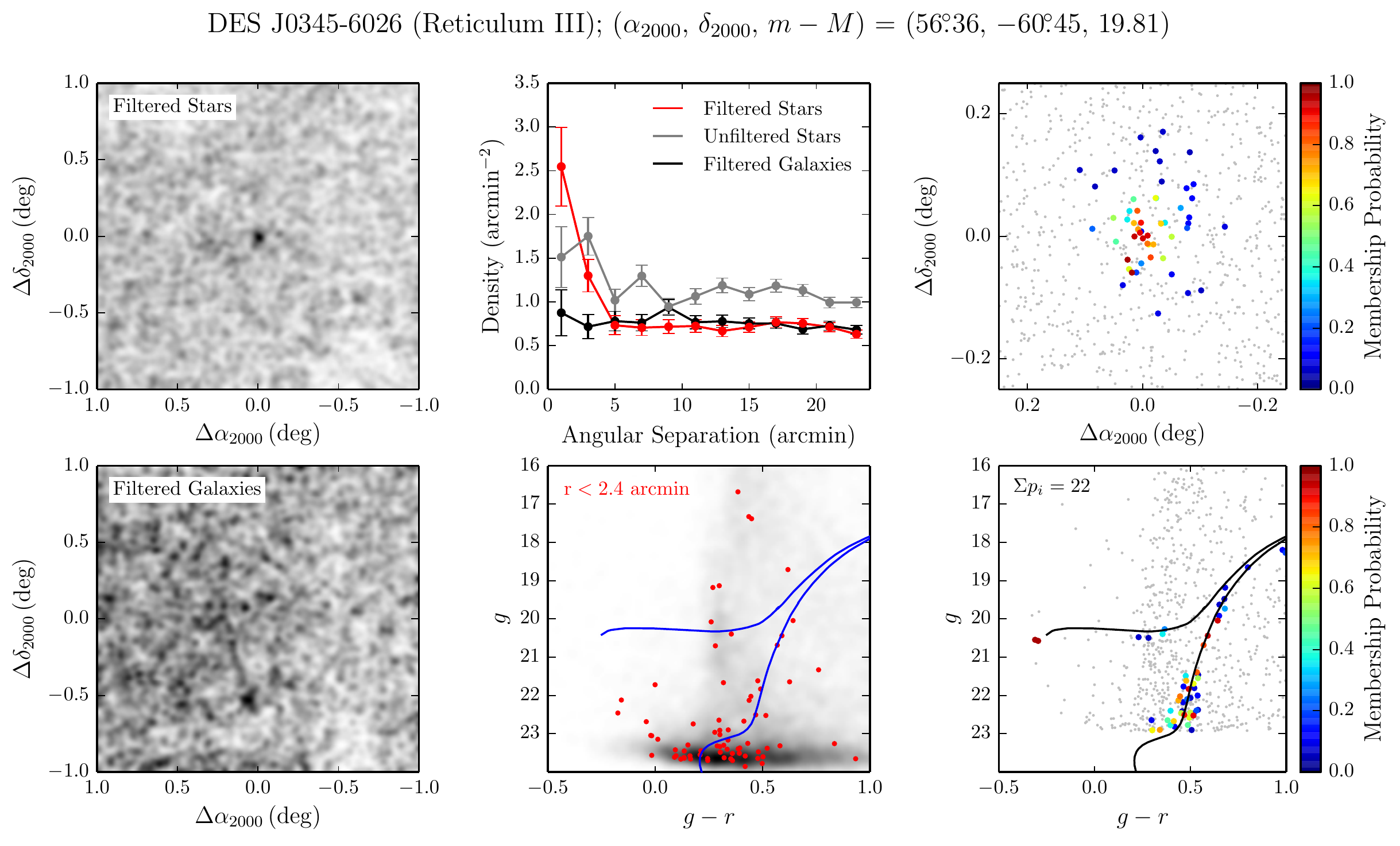}
  \caption{Analogous to \figref{cmd_gruII} but for \retIII (Reticulum~III).
  }
\label{fig:cmd_retIII}
\end{figure*}

\item {\bf \tucV} (Tucana~V, \figref{cmd_tucV}): 
\tucV is the second new system in the Tucana group and the closest to the group's centroid (\CHECK{$3 \kpc$}).
At a heliocentric distance of \CHECK{$55 \kpc$}, \tucV is also located \CHECK{$29\kpc$} from the LMC and \CHECK{$14\kpc$} from the SMC.
The physical size (\CHECK{$17 \pc$}) and luminosity (\CHECK{$M_V \sim -1.6 \magn$}) place \tucV in a region of the size-luminosity plane close to Segue~1, Willman~1, and Kim~2.
Segue~1 and Willman~1 have sizes similar to the most extended globular clusters, but are approximately an order of magnitude less luminous.
On the other hand, deep imaging of Kim~2 has led \citet{2015ApJ...803...63K} to conclude that it is very likely a star cluster.
\tucV is one of the few new objects for which the DES data places a constraint on ellipticity.
The high ellipticity, \CHECK{$\epsilon = 0.7$}, supports a galactic classification for \tucV.
Most of the high probability member stars for \tucV are on the main sequence and fainter than $g\sim22.5 \magn$.
As with other similarly faint objects, a few stars situated close to the lower RGB are also likely members.
A system similar to \tucV would have been difficult to detect in SDSS with a threshold at $r < 22$.
\begin{figure*}
  \includegraphics[width=1.\textwidth]{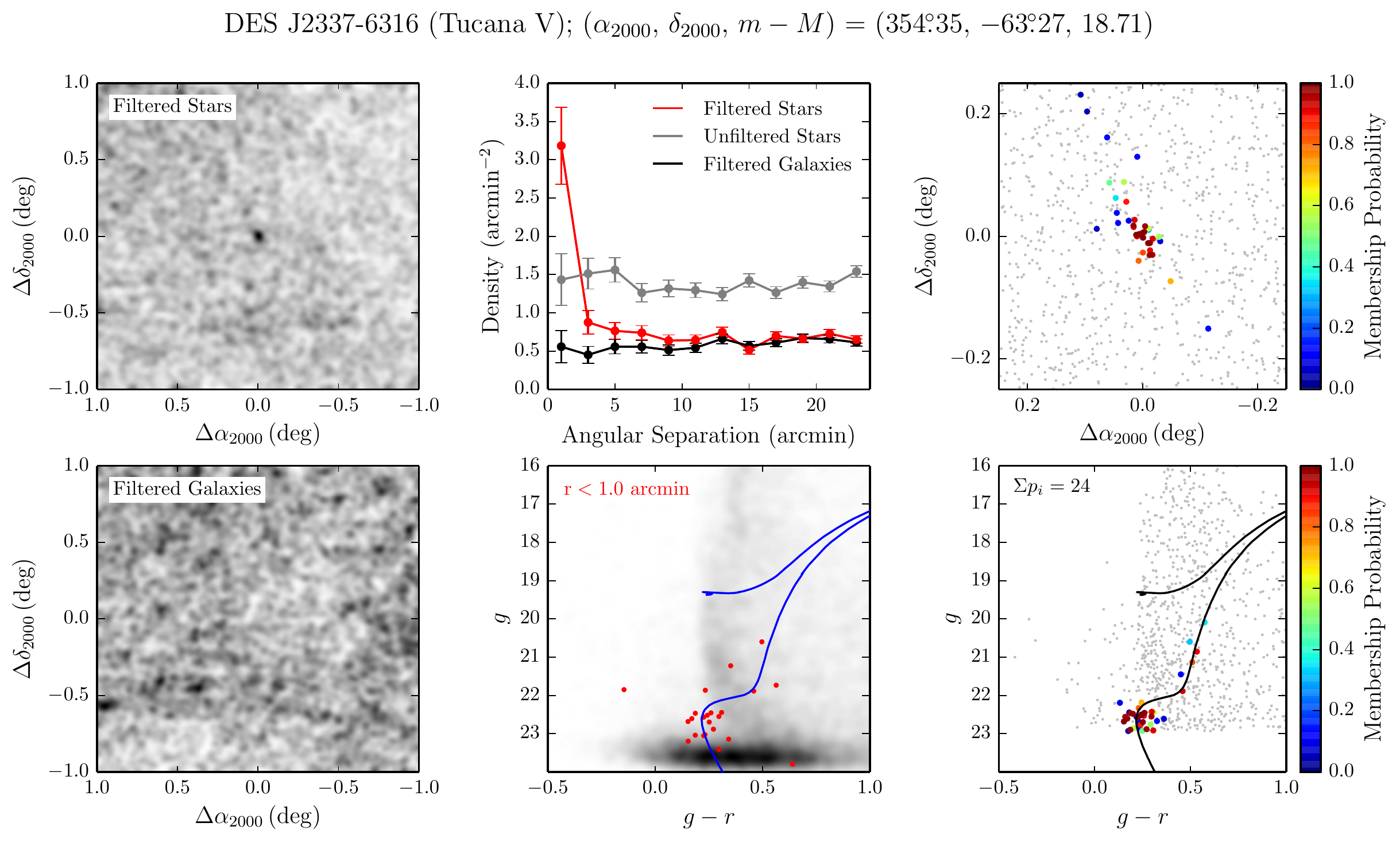}
  \caption{Analogous to \figref{cmd_gruII} but for \tucV (Tucana~V).
  }
\label{fig:cmd_tucV}
\end{figure*}

\item {\bf \indII} (Indus~II, \figref{cmd_indI}):
\indII is the first of the two lower confidence candidates.
The detection of this object comes predominantly from a tight clump of BHB stars at $g \sim 22 \magn$ and $g-r \sim 0 \magn$.
Three of the potential HB members are clustered within a spatial region of radius $\roughly 10 \arcsec$.
Several additional HB and RGB stars are assigned non-zero membership probabilities by the likelihood fit, enlarging the best-fit size of this system.
\indII is located at a boundary between the Y1 and Y2 imaging, and the effect of the deeper Y2 source detection threshold is visible when extending to magnitudes of $g > 23 \magn$. 
The best-fit physical size of \indII (\CHECK{$181\pc$}) would place it among the population of Local Group galaxies.
Deeper imaging is needed to better characterize this candidate since the MSTO is fainter than the current DES detection threshold.

\begin{figure*}
  \includegraphics[width=1.\textwidth]{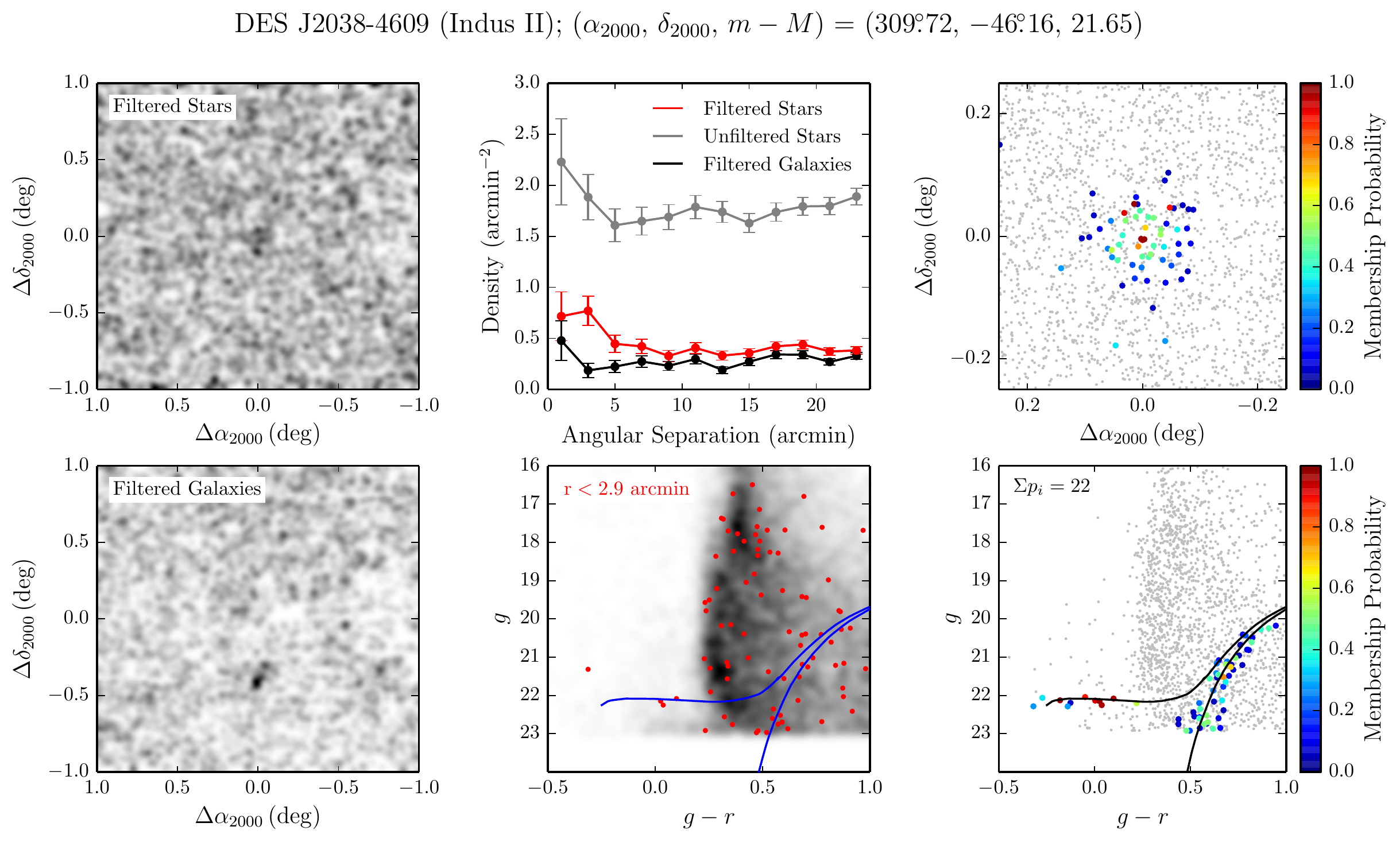}
  \caption{Analogous to \figref{cmd_gruII} but for \indII (Indus~II).
  Note that the magnitude threshold for the smoothed counts maps and radial profiles has been raised to $g < 23 \magn$ due to non-uniform imaging depth in this region.
  \indII is detected primarily by a cluster of potential BHB members.
  }
\label{fig:cmd_indI}
\end{figure*}

\item {\bf \cetII} (Cetus~II, \figref{cmd_cetII}): 
\cetII is the second lower confidence candidate, being the least luminous (\CHECK{$M_V = 0.0 \pm 0.68 \magn$}) and most compact (\CHECK{$r_{1/2} = 17\pc$}) of the new stellar systems. 
\cetII is nearly two orders of magnitude fainter than globular clusters with comparable half-light radii.
At a heliocentric distance of \CHECK{$30\kpc$}, \cetII is the second nearest of the new systems.
\cetII is detected predominantly by its main sequence stars and has one potential HB member.
The lack of strong features in the color-magnitude distribution of stars associated with \cetII leads to a large degeneracy between the distance, age, and metallicity of this system.
If determined to be a dwarf galaxy, \cetII would be the least luminous known galaxy; however, the current classification of this object is ambiguous.

There are several gaps in the Y2Q1 coverage $\sim 0\fdg1$ from \cetII; however, \cetII is sufficiently compact that few member stars are expected in the region of missing coverage.
These gaps are incorporated into the likelihood analysis and should have a minimal impact on the fit.
The region around \cetII was imaged in only a single $r$-band exposure; however, there are two to three overlapping exposures in each of the $g$- and $i$-bands.
We have verified that the stellar overdensity is also apparent in the $g-i$ filter combination.
Nonetheless, the properties of \cetII should be interpreted with caution until additional imaging is acquired.

\begin{figure*}
  \includegraphics[width=1.\textwidth]{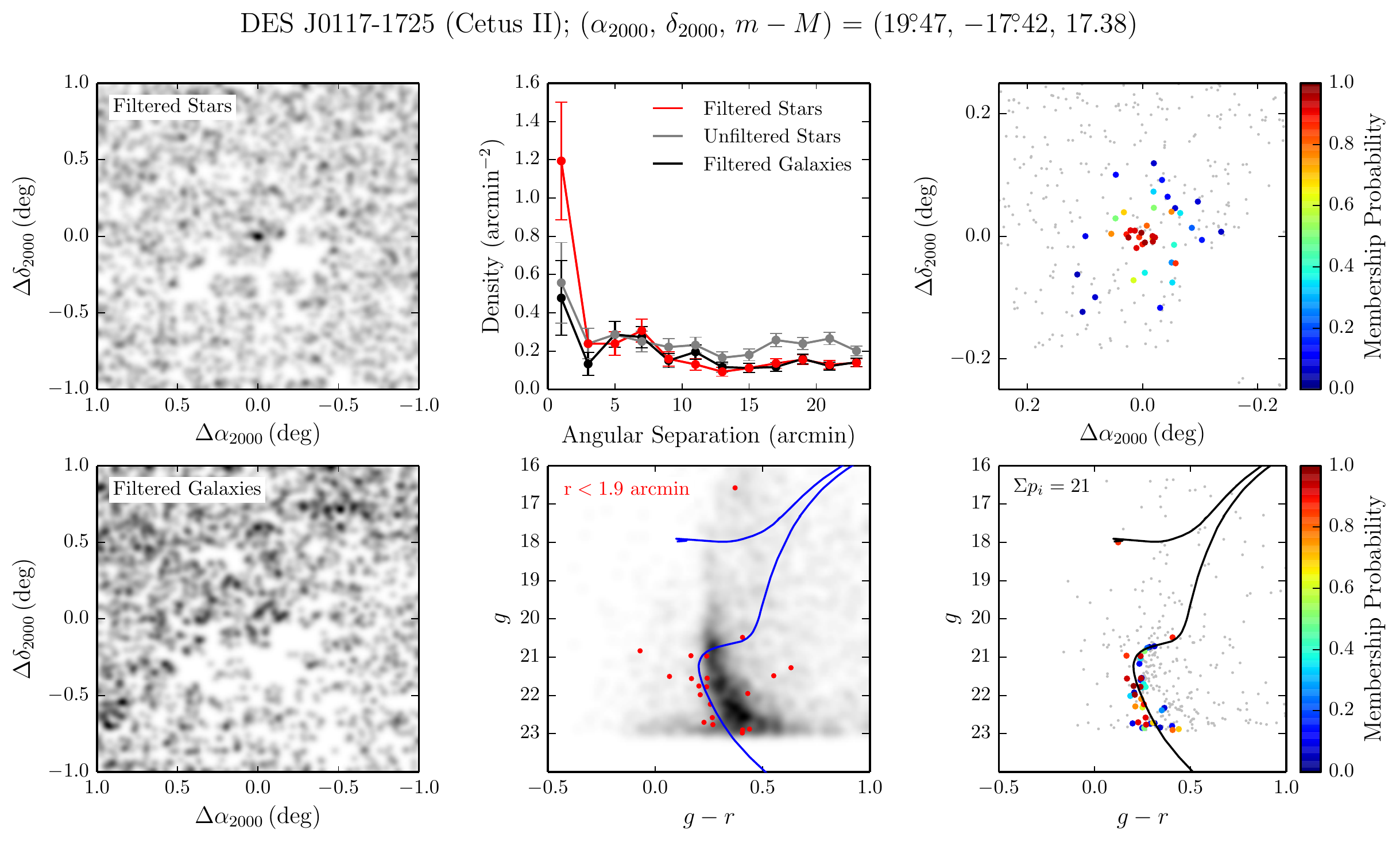}
  \caption{Analogous to \figref{cmd_gruII} but for \cetII (Cetus~II). 
  Note that the magnitude threshold for the smoothed counts maps and radial profiles has been raised to $g < 23 \magn$ due to incomplete coverage in this region. }
\label{fig:cmd_cetII}
\end{figure*}

\end{itemize}

\subsection{Satellite Detection Completeness}
\label{sec:completeness}

Many studies of the Milky Way satellite galaxy population require knowledge of the detection efficiency for satellites as a function of luminosity, heliocentric distance, physical size, and sky position in order to account for observational selection effects.
To quantify our search sensitivity with the Y2Q1 data set, we follow the approach that \citet{Bechtol:2015wya} applied to the DES Y1A1 coadd data.
Briefly, we generated many realizations of satellite galaxies with different structural properties that are spread uniformly throughout the survey footprint, excluding regions of high stellar density near to the LMC.
We then applied the map-based search algorithm to estimate the likelihood of a high-confidence detection.
In the present implementation, the detection significance was evaluated as the Poisson probability of finding the ``observed'' number of stars within a circular region matched to the half-light radius of the simulated satellite given the effective local field density after the isochrone selection (see \secref{stellar_density_maps}).
Realizations that yield at least 10 detectable stars and produce a $> 5.5 \sigma$ overdensity within the half-light radius are considered ``detected''.

An essential input to the completeness calculation is the magnitude threshold for individual stars.
For the searches described in \secref{methods}, we used deeper magnitude thresholds (\eg, $g < 24 \magn$) in order to maximize the discovery potential.
A more conservative threshold, $g < 23 \magn$, is selected to estimate the completeness in detecting new stellar systems based on the following considerations.
First, the Y2Q1 stellar sample is expected to be nearly complete for both Y1 and Y2 exposures at magnitudes $g \lesssim 23 \magn$.
\figref{skymap_zoom} shows that the measured stellar density in this magnitude range varies smoothly over the Y2Q1 footprint.
At fainter magnitudes, discontinuities in the field density can occur in regions where exposures from Y1 and Y2 overlap (typically, more faint objects are detected in the Y2 exposures).
Second, the field population with $g \gtrsim 23 \magn$ in high Galactic latitude regions is likely dominated by misclassified galaxies rather than halo stars, and therefore the assumption of isotropy on arcminute angular scales is compromised due to galaxy clustering.
The adopted magnitude threshold matches that used in \citet{Bechtol:2015wya}, and in this case, the sensitivity estimates obtained in the prior work can be approximately scaled to the enlarged effective solid angle of the Y2Q1 data set, $\roughly 5000 \deg^2$.

In \tabref{efficiency}, we list the expected detection efficiencies for the DES ultra-faint galaxy candidates using magnitude thresholds of $g < 23 \magn$ and $g < 22 \magn$, which roughly correspond to the stellar completeness limits of Y2Q1 and SDSS, respectively.
For the brighter threshold, several of the DES candidates have substantially reduced detection probabilities.

We note that all of the faint stellar systems that were discovered in the Y1A1 data set are also significantly detected using the Y2Q1 data set, suggesting that our present search has sensitivity comparable to previous studies.

\newcommand{\effcaption}{Expected detection efficiencies for ultra-faint galaxy candidates.\label{tab:efficiency}}
\newcommand{\effcomments}{
Expected detection efficiencies are provided for two magnitude thresholds roughly corresponding to the stellar completeness depths of DES Y2Q1 ($g < 23 \magn$) and SDSS ($g < 22 \magn$).
The detection efficiencies are estimated from multiple realizations of each candidate within the DES footprint using the best-fit luminosity $M_V$, heliocentric distance, and azimuthally averaged half-light radius $r_h$, which were then analyzed with the map-based detection algorithm described in \secref{stellar_density_maps}.
The average detection probability ratio between the two threshold choices is $\roughly 50 \%$, implying that roughly half of the DES candidates would have been detected if they were located in the SDSS footprint.}
\begin{\tabletype}{l ccccc }
\tablecolumns{6}
\tablewidth{0pt}
\tabletypesize{\tiny}
\tablecaption{ \effcaption }
\tablehead{
Name & $M_V$ & Distance & $r_h$    & Efficiency                       & Efficiency \\ 
     &       & (kpc)    & (arcmin) & ($g < 23 \magn$) & ($g < 22 \magn$)
}
\startdata
\gruII (Gru\,II) & -3.9 & 53 & 6.0 & 1.00 & 0.39 \\
\tucIII (Tuc\,III) & -2.4 & 25 & 6.0 & 1.00 & 0.94 \\
\colI (Col\,I) & -4.5 & 182 & 1.9 & 0.95 & 0.86 \\
\tucIV (Tuc\,IV) & -3.5 & 48 & 9.1 & 0.98 & 0.05 \\
\retIII (Ret\,III) & -3.3 & 92 & 2.4 & 0.33 & 0.05 \\
\tucV (Tuc\,V) & -1.6 & 55 & 1.0 & 0.94 & 0.01 \\
\indII (Ind\,II) & -4.3 & 214 & 2.9 & 0.26 & 0.00 \\
\cetII (Cet\,II) & 0.0 & 30 & 1.9 & 0.38 & 0.01 \\
\vspace{-0.2cm}\\\tableline\tableline\vspace{-0.2cm}\\
Ret\,II & -3.6 & 32 & 6.0 & 1.00 & 1.00 \\
\eriII (Eri\,II) & -7.4 & 330 & 1.8 & 1.00 & 1.00 \\
\tucII (Tuc\,II) & -3.9 & 58 & 7.2 & 1.00 & 0.07 \\
Hor\,I & -3.5 & 87 & 2.4 & 0.66 & 0.14 \\
\picI (Pic\,I) & -3.7 & 126 & 1.2 & 0.85 & 0.47 \\
\eriIII (Eri\,III) & -2.4 & 95 & 0.4 & 0.41 & 0.04 \\
\pheII (Phe\,II) & -3.7 & 95 & 1.2 & 0.99 & 0.77 \\
Gru\,I & -3.4 & 120 & 2.0 & 0.21 & 0.04 \\
Hor\,II & -2.6 & 78 & 2.1 & 0.26 & 0.01 \\
\enddata
{\footnotesize \tablecomments{ \effcomments }}
\end{\tabletype}

\subsection{Total Number and Distribution of Milky Way Satellite Galaxies}
\label{sec:distribution}

\begin{figure}
\centering
\includegraphics[width=0.75\textwidth]{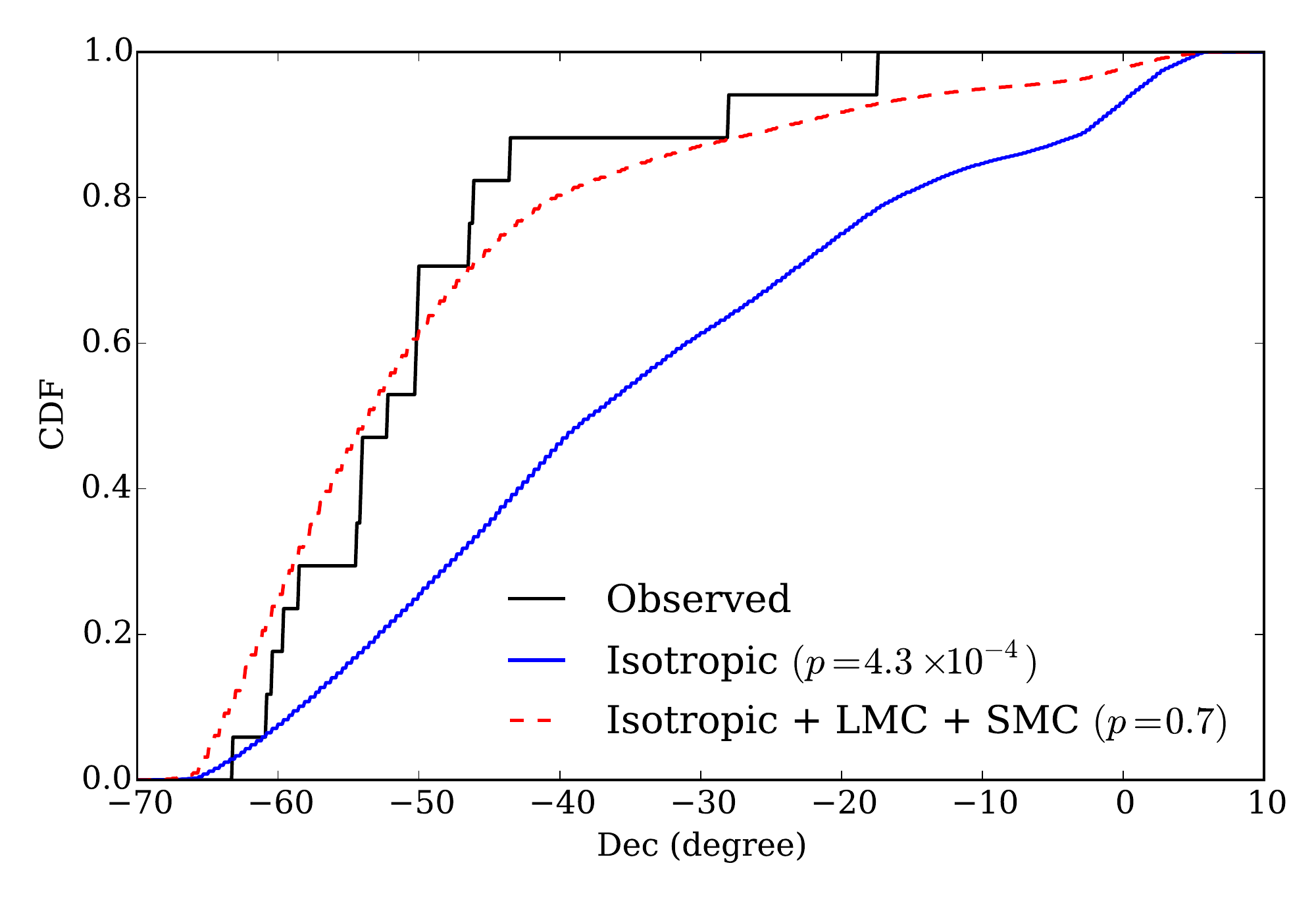}
\caption{The cumulative distribution functions of declination values for the \ndes observed satellites in the DES footprint (black), a uniform distribution given the shape of DES footprint (blue), and the best-fit model for a uniform distribution and Magellanic Cloud components (red dashed). See \secref{distribution}.}
\label{fig:kstest}
\end{figure}

\begin{figure*}
\centering
\includegraphics[width=\textwidth]{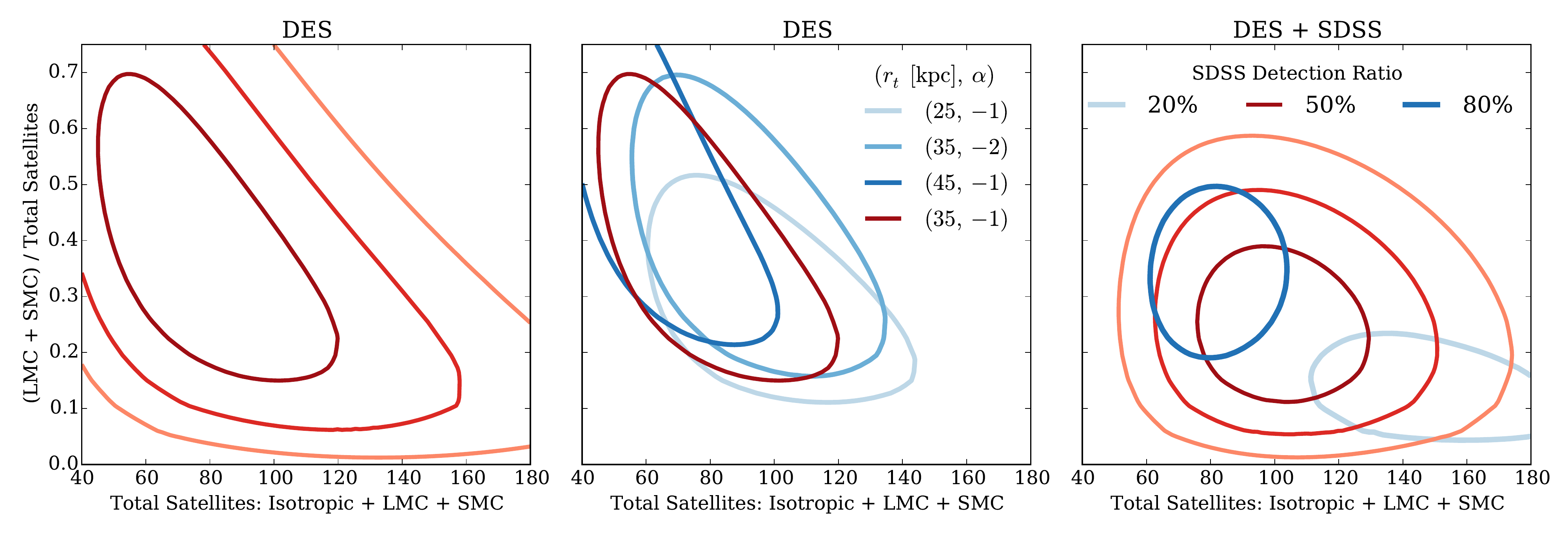}
\caption{ Maximum-likelihood fit to the spatial distribution of satellites for a model including isotropic, LMC, and SMC components (\secref{distribution}).
The horizontal axis represents the total number of satellites detectable at DES Y2Q1 depth integrated over the entire sky (not including the classical dwarf galaxies and detection inefficiency near the Galactic plane).
The vertical axis represents the fraction of these satellites associated with the Magellanic Clouds.
\textit{Left}: The $1\sigma$, $2\sigma$, and $3\sigma$ likelihood contours when considering only DES data.
\textit{Center}: $1\sigma$ contours for different values of the LMC and SMC truncation radius ($r_t$) and the slope of the radial profile ($\alpha$).
\textit{Right}: Contours of the likelihood function when SDSS observations are included. Red contours ($1\sigma$, $2\sigma$, $3\sigma$) assume that 50\% of the satellites discovered in DES would have been detected if they were located in the SDSS footprint. 
This value matches the estimated detection efficiency from \tabref{efficiency}. 
We show the sensitivity of our results to the SDSS/DES detection ratio with the dark blue contour (80\% detection ratio) and the light blue contour (20\% detection ratio).
}
\label{fig:likelihood}
\end{figure*}

%\FIXME{The assumptions regarding isotropy/anisotropy of the Milky Way satellite population for the references in first paragraph should be more clearly explained in the revised version.}

The detection of \ndes candidate ultra-faint galaxies in the first two years of DES data is consistent with the range of predictions based on the standard cosmological model \citep{2008ApJ...688..277T,2014ApJ...795L..13H}.
These predictions model the spatial distribution of luminous satellites from the locations of dark matter subhalos in cosmological $N$-body simulations of Local Group analogs.
Significant anisotropy in the distribution of satellites is expected \citep{2008ApJ...688..277T,2014ApJ...795L..13H}, and likely depends on the specific accretion history of the Milky Way \citep{Deason:2015b}.

The proximity of the DES footprint to the Magellanic Clouds is noteworthy when considering an anisotropic distribution of satellites.
The possibility that some Milky Way satellites are or were associated with the Magellanic system has been discussed for some time \citep{1976MNRAS.174..695L,D'Onghia:2008a,Sales:2011a,Nichols:2011a}, and the recent DES results have renewed interest in this topic \citep{Koposov:2015cua,Deason:2015b,Wheeler:2015,Yozin:2015}.  
The existence of structures on mass scales ranging from galaxy clusters to ultra-faint satellite galaxies is a generic prediction of hierarchical galaxy formation in the cold dark matter paradigm.  
However, specific predictions for the number of observable satellites of the LMC and SMC depend on the still unknown efficiency of galaxy formation in low-mass subhalos.  
Identifying satellites of satellites within the Local Group would support the standard cosmological model, and may provide a new opportunity to test the impact of environment on the formation of the least luminous galaxies \citep{Wetzel:2015a}.

Based on a study of Milky Way--LMC analogs in $N$-body simulations, \citet{Deason:2015b} propose a method to evaluate the probability that a given satellite was at one time associated with the LMC based upon the present three-dimensional separation between the two (\tabref{separations}).
This prescription predicts that two to four of the Y2 DES candidates were associated with the LMC before infall, in addition to the two to four potentially associated satellites found in Y1 DES data.
\citeauthor{Deason:2015b} further suggest that a grouping of satellites near to the LMC in both physical separation and velocity would imply that the Magellanic system was recently accreted onto the Milky Way since members of this group would disperse within a few Gyr.

As a population, the new DES satellite galaxy candidates do appear to be unevenly distributed within the survey footprint; \CHECK{fifteen} are located at $\delta_{2000} < -40^{\circ}$, close to the Magellanic Clouds (\figref{skymap_zoom}).
A Kolmogorov-Smirnov test, shown in \figref{kstest}, significantly rejects the hypothesis of a uniform distribution ($p = 4.3 \times 10^{-4}$).  
One possible explanation for non-uniformity is that some of the new systems are satellites of the LMC or SMC.  
Below we introduce a simple model to test this hypothesis.

We began by modeling the probability distribution for detecting a satellite galaxy as a function of sky position.  
In this analysis, we excluded classical Milky Way satellites and considered only satellites that have similar characteristics to those discovered by DES (\ie, are detectable at the DES Y2Q1 depth).  
The underlying luminosity function and the completeness as a function of luminosity, heliocentric distance, and physical size were all simplified into a two-dimensional probability distribution.

We modeled this two-dimensional probability distribution with three components: ($i$) an isotropic distribution of satellites associated with the main halo of the Milky Way, ($ii$) a population of satellites spatially associated with the LMC halo, and ($iii$) a population of satellites spatially associated with the SMC halo.  
The first component was uniform on the sky, and the latter two components were spatially concentrated around the Magellanic Clouds.  
We assumed spherically symmetric three-dimensional distributions of satellites centered around each of the Magellanic Clouds having a power-law radial profiles with slope $\alpha = -1$ and truncation radius $r_t = 35\kpc$. 
These distributions were then projected onto the sky to create two-dimensional predicted density maps.

Several underlying assumptions went into this simple model.  
First, we assumed that the Magellanic components were concentrated enough that variations in the completeness with respect to heliocentric distance would not substantially alter the projected density maps. 
Second, we assumed that the truncation radii of the LMC and SMC are the same, but later test the sensitivity of our results to this choice.  
Third, we assumed that smaller satellites of the Milky Way do not have their own associated satellites.  
We note that a power-law radial profile with slope $\alpha = -1$ can be viewed as a very extended Navarro-Frenk-White profile \citep{Navarro:1997}.
Since the DES footprint does not cover the central regions of the Magellanic Clouds, this shallow profile yields a conservative estimate on the fraction of satellites associated with the Magellanic system.

We performed an unbinned maximum-likelihood fit of our three-component model to the distribution of observed satellites. 
The parameters of the likelihood function are the total number of satellites and the relative normalizations of the three components.
The likelihood is defined as the product of the probabilities to detect each observed satellite given its sky position. 
When presenting our results, we marginalized over the ratio of LMC component to SMC component. 
This ratio is not strongly constrained by our analysis; however, there is a slight preference for a larger SMC component.

The left panel of \figref{likelihood} shows the constraints using the sky positions of the \ndes DES satellite galaxy candidates.  
We find that the DES data alone reject a uniform distribution of satellites at $>3\sigma$ confidence when compared to our best-fit model.  
This again implies that there is either a clear overdensity of satellites around the Magellanic Clouds, or that the initial assumption of isotropy around the main halo is incorrect.  
In \figref{kstest}, we show the distribution of declinations for our best-fit three-component model and find good agreement with the observations ($p = 0.7$).

We next tested the sensitivity of our result to the assumed distribution of satellites around the LMC and SMC.
The central panel of \figref{likelihood} shows the $1\sigma$ contours for three different values of $r_t = \{25, 35, 45 \kpc\}$, and two values of the slope, $\alpha = \{-1, -2\}$. 
Despite slight changes in the likelihood contours, in each case $>10\%$ of Milky Way satellites are likely to be spatially associated with the Magellanic Clouds.

We further examined the satellite population by simultaneously considering the \nsdss ultra-faint satellites observed in the SDSS DR10 footprint.\footnote{We include Pegasus~III \citep{kim_2015_pegasus_iii} in the list of ultra-faint satellites from SDSS.}
To fully combine DES and SDSS observations we would need to model the detection efficiency of SDSS relative to DES as a function of satellite distance, size, and luminosity.  
However, since the SDSS DR10 footprint only overlaps with the isotropic component in our model, we can again fold the complications of detection efficiency into a constant detection ratio.  
Our fiducial calculation assumes that 50\% of the satellites discovered in DES would be detectable by SDSS, which agrees with the relative detection efficiency of the \ndes DES satellites at the depth of SDSS (\tabref{efficiency}).
The right panel of \figref{likelihood} shows that the DES+SDSS constraints are in agreement with those from DES data alone, though the fraction of LMC and SMC satellites is more tightly constrained by including information from a large region widely separated from the Magellanic Clouds.
The combined results imply that there are $\roughly100$ satellites over the full sky that are detectable at DES Y2Q1 depth, and that 20--30\% of these might be associated with the Magellanic Clouds.
Importantly, this prediction does not attempt to model the full population of Milky Way satellites beyond the DES Y2Q1 sensitivity and ignores the diminished detection efficiency close to the Galactic plane.

The Pan-STARRS team has recently identified three candidate Milky Way satellite galaxies in their 3$\pi$ survey \citep{Laevens:2015a,Laevens:2015b}. 
If this search is complete to the depth of SDSS over its full area, it is notable that so few candidates have been found.
This observation alone could imply anisotropy without requiring the influence of the Magellanic Clouds.  
However, much of the Pan-STARRS area is located at low Galactic latitudes where the elevated foreground stellar density and interstellar extinction may present additional challenges.
Under the assumption that Pan-STARRS covers the full sky with $\dec > -30^{\circ}$, $\roughly 2000 \deg^2$ overlap with DES Y2Q1. 
This area of the DES footprint includes two new candidates, one of which has a large enough surface brightness to likely have been detected at SDSS depth (\colI, see \tabref{efficiency}).
\CHECK{These two candidates are located in a region of the sky that would be observed at a relatively high airmass by Pan-STARRS and may suffer from decreased detection efficiency.}

From this analysis, we conclude that the distribution of satellites around the Milky Way is unlikely to be isotropic, and that a plausible component of this anisotropy is a population of satellites associated with the Magellanic Clouds.  
However, several alternative explanations for anisotropy in the Milky Way satellite distribution exist.  
For example, Milky Way satellites could be preferentially located along a three-dimensional planar structure, as has been suggested by many authors starting with \cite{1976MNRAS.174..695L}.  
This proposed planar structure encompasses the Magellanic Clouds and many of the classical and SDSS satellites.
\citet{Pawlowski:2015dua} suggest that the satellites discovered in Y1 DES data are also well aligned with this polar structure. 
We note that the Y2 discoveries presented here include several objects near the SMC and may reduce the fraction of objects in close proximity to the proposed plane. 
An additional possibility is that the satellites are associated with the orbit of the Magellanic System and are not isotropically distributed around the Magellanic Clouds themselves \citep[][]{Yozin:2015}.  
The DES footprint covers only a fraction of the region surrounding the Magellanic Clouds and additional sky coverage may yield more satellites with similar proximity to the Magellanic system and/or help to distinguish between these various scenarios.  
Measurements of the relative motions of the satellites and further theoretical work will also help clarify the physical relationships between these stellar systems.

\section{Conclusions}
\label{sec:conclusions}

We report the discovery of \nobjs new ultra-faint galaxy candidates in a combined data set from the first two years of DES covering $\roughly 5000 \deg^2$ of the south Galactic cap.
\Ncand additional candidates are identified in regions with incomplete or non-uniform coverage and should be viewed with lower confidence until additional imaging is obtained.
The new satellites are faint ($M_V > \lummax \magn$) and span a wide range of physical sizes ($\sizemin \pc < r_{1/2} < \sizemax \pc$) and heliocentric distances ($\distmin\kpc < D_\odot < \distmax\kpc$).
All are low surface brightness systems similar to the known ultra-faint satellite galaxies of the Milky Way, and most possess physical sizes that are large enough ($>40\pc$) to be provisionally classified as galaxies. 
%\cetII and \tucV lie closer to the globular cluster population in size-luminosity space.
% that may be indicative of a gravitation interaction with the Milky Way.
Spectroscopic observations are needed to better understand and unambiguously classify the new stellar systems.
A total of \ndes confirmed and candidate ultra-faint galaxies have been found in the first two years of DES.
Roughly half of the DES systems are sufficiently distant and/or faint to have eluded detection at survey depths comparable to SDSS.

The DES satellites are concentrated in the southern half of the survey footprint in proximity to the Magellanic Clouds.  
In addition, we find three satellites clustered in a Tucana group, each of which is within $< 10 \kpc$ of the group centroid.
We find that the DES data alone exclude ($p < 10^{-3}$) an isotropic distribution of satellites within the Milky Way halo, and that the observed distribution can be well, although not uniquely, explained by a model in which several of the observed DES satellites are associated with the Magellanic system.  
Under the assumption that the total satellite population can be modeled by isotropic distributions around the Milky Way, LMC, and SMC, we estimate that a total of $\roughly100$ ultra-faint satellites with comparable physical characteristics to those detected by DES might exist over the full sky, with 20--30\% of these systems being spatially associated with the Magellanic Clouds.

Milky Way satellite galaxies are considered a unique population for studying the particle nature of dark matter due to their proximity, characteristically large mass-to-light ratios, and lack of intrinsic astrophysical backgrounds.
Although the dark matter content of the new stellar systems has not yet been spectroscopically confirmed, the possibility that some are dark matter dominated galaxies makes them interesting targets for indirect dark matter searches via gamma rays.
No gamma-ray sources from the third \textit{Fermi}-LAT source catalog \citep[3FGL;][]{2015ApJS..218...23A} are located within \CHECK{0\fdg5} of any of the new candidates.
A follow-up gamma-ray analysis of these ultra-faint galaxy candidates will be presented separately.

Future seasons of DES will not significantly increase the DES sky coverage. 
However, there will be considerable gains in depth from the expected ten tilings per filter relative to the two to four in the Y2Q1 data set.
% (relative to the two to four tilings for Y2Q1).
This increased depth will expand the effective volume of the survey.
In addition, we expect the survey uniformity to improve with the coming seasons, resulting in a cleaner list of seeds with fewer false positives.
Star-galaxy separation may become a limiting factor at the future depth of DES, motivating the development of improved classification algorithms \citep[\eg,][]{2012ApJ...760...15F,2013arXiv1306.5236S}, as well as alternative search strategies, involving, for instance, the time domain \citep{Baker:2015} or stellar velocities \citep{Antoja:2015}.
While it is likely that the most conspicuous Milky Way satellites in the DES footprint have been found, the most exciting phase of Milky Way science with DES is likely still to come.
The wide area and growing sensitivity of DES will enable the discovery of dwarf galaxies that are fainter, farther, and lower surface brightness.

\section{Acknowledgments}

This work made use of computational resources at SLAC National Accelerator Laboratory and the University of Chicago Research Computing Center.
Some of the results in this paper have been derived using the \HEALPix \citep{2005ApJ...622..759G} package.
This research made use of Astropy, a community-developed core Python package for Astronomy \citep{2013A&A...558A..33A}.
We thank the anonymous referee for helpful suggestions.
ADW thanks Ellen Bechtol for her generous hospitality during the preparation of this manuscript.
EB acknowledges financial support from the European Research Council (ERC-StG-335936, CLUSTERS).

Funding for the DES Projects has been provided by the U.S. Department of Energy, the U.S. National Science Foundation, the Ministry of Science and Education of Spain, 
the Science and Technology Facilities Council of the United Kingdom, the Higher Education Funding Council for England, the National Center for Supercomputing 
Applications at the University of Illinois at Urbana-Champaign, the Kavli Institute of Cosmological Physics at the University of Chicago, 
the Center for Cosmology and Astro-Particle Physics at the Ohio State University,
the Mitchell Institute for Fundamental Physics and Astronomy at Texas A\&M University, Financiadora de Estudos e Projetos, 
Funda{\c c}{\~a}o Carlos Chagas Filho de Amparo {\`a} Pesquisa do Estado do Rio de Janeiro, Conselho Nacional de Desenvolvimento Cient{\'i}fico e Tecnol{\'o}gico and 
the Minist{\'e}rio da Ci{\^e}ncia, Tecnologia e Inova{\c c}{\~a}o, the Deutsche Forschungsgemeinschaft and the Collaborating Institutions in the Dark Energy Survey. 

The Collaborating Institutions are Argonne National Laboratory, the University of California at Santa Cruz, the University of Cambridge, Centro de Investigaciones Energ{\'e}ticas, 
Medioambientales y Tecnol{\'o}gicas-Madrid, the University of Chicago, University College London, the DES-Brazil Consortium, the University of Edinburgh, 
the Eidgen{\"o}ssische Technische Hochschule (ETH) Z{\"u}rich, 
Fermi National Accelerator Laboratory, the University of Illinois at Urbana-Champaign, the Institut de Ci{\`e}ncies de l'Espai (IEEC/CSIC), 
the Institut de F{\'i}sica d'Altes Energies, Lawrence Berkeley National Laboratory, the Ludwig-Maximilians Universit{\"a}t M{\"u}nchen and the associated Excellence Cluster Universe, 
the University of Michigan, the National Optical Astronomy Observatory, the University of Nottingham, The Ohio State University, the University of Pennsylvania, the University of Portsmouth, 
SLAC National Accelerator Laboratory, Stanford University, the University of Sussex, and Texas A\&M University.

The DES data management system is supported by the National Science Foundation under Grant Number AST-1138766.
The DES participants from Spanish institutions are partially supported by MINECO under grants AYA2012-39559, ESP2013-48274, FPA2013-47986, and Centro de Excelencia Severo Ochoa SEV-2012-0234.
Research leading to these results has received funding from the European Research Council under the European Union’s Seventh Framework Programme (FP7/2007-2013) including ERC grant agreements  240672, 291329, and 306478.

\bibliographystyle{apj}
\bibliography{main}

\end{document}